\documentclass{llncs}
\usepackage{orcidlink}

\title{Satisfiability of
  Non-Linear Transcendental Arithmetic as a
  Certificate Search Problem}

\usepackage{url}
\usepackage{todonotes}
\usepackage{hyperref}

\usepackage{tableof}

\usepackage{xcolor}
\usepackage{amsmath}
\usepackage{amsfonts}
\usepackage{graphicx}
\usepackage{xspace}
\usepackage{hhline}
\usepackage{makecell}
\usepackage{colortbl} 

\usepackage{amssymb}

\newcommand{\nta}{\ensuremath{\mathcal{N\mkern-1mu T\mkern-4mu A}}\xspace}

\newcommand{\smtnta}{\ensuremath{\text{SMT}(\nta)}\xspace}

\newcommand{\ugotNL}{\textsc{ugotNL}\xspace}
\newcommand{\ugotNLeager}{$\ugotNL_{\text{\textsc{eager}}}$\xspace}

\newcommand{\mathsat}{\textsc{MathSAT}\xspace}
\newcommand{\cvctool}{\textsc{cvc5}\xspace}
\newcommand{\drealtool}{\textsc{dReal}\xspace}
\newcommand{\isattool}{\textsc{iSAT3}\xspace}
\newcommand{\msatUgot}{\textsc{MathSAT+ugotNL}\xspace}

\newcommand{\myparagraph}[1]{\ \\\textbf{#1}}
\newcommand{\myparagraphB}[1]{\emph{\normalsize{#1}}}

\newcommand{\Sat}{\texttt{sat}}
\newcommand{\Unknown}{\texttt{unknown}}
\newcommand{\Unsat}{\texttt{unsat}}
\newcommand{\Vars}[1]{\mathit{vars}_{\mathcal{R}}(#1)}
\newcommand{\LtoO}{\ensuremath{\mathcal{L}2\mathcal{O}}}
\newcommand{\defas}{\ensuremath{\stackrel{\text{\tiny def}}{=}}\xspace}

\newcommand{\R}{\ensuremath{\mathbb{R}}\xspace}

\newcommand{\openball}[2]{\ensuremath{\mathcal{#1}_{#2}}}

\newcommand{\kfinite}[1]{\ensuremath{{#1}}-finite point\xspace}
\newcommand{\kfinites}[1]{\ensuremath{{#1}}-finite points\xspace}

\newcommand{\FRob}{\mathcal{F}_{Rob}\xspace}

\newcommand{\FRobI}{\mathcal{F}_{RobI}\xspace}
\newcommand{\FReg}{\mathcal{F}_{Reg}\xspace}
\newcommand{\FRobLEq}{\mathcal{F}_{RobLEq}\xspace}

\ifx\hideTODO \undefined

\newcommand\SRtodo[1]{\todo[backgroundcolor=cyan]{#1}}
\newcommand\ELtodo[1]{\todo[backgroundcolor=yellow,]{#1}}

\else
\newcommand\SRtodo[1]{}
\newcommand\ELtodo[1]{}
\fi

\newcommand{\heuristic}[1]{\texttt{(#1)}\xspace}

\newcommand{\filterOverconstr}{\heuristic{filter-overconstr}}
\newcommand{\sortWrtCost}{\heuristic{sort-literals}}
\newcommand{\checkForcedLiterals}{\heuristic{check-forced-literals}}

\newcommand{\filterOverconstrV}{\heuristic{filter-overconstr-V}}
\newcommand{\filterRankDeficient}{\heuristic{filter-rank-deficient}}
\newcommand{\KearfottOrdering}{\heuristic{Kearfott-ordering}}

\newcommand{\epsInflation}{\heuristic{eps-inflation}}
\newcommand{\boxGridding}{\heuristic{box-gridding}}

\newcommand{\orthogonal}{\heuristic{orthogonal}}
\newcommand{\gaussel}{\heuristic{gauss-elim}}

\newcommand{\expid}[2]{\ ({#1}.{#2}.) \ \ }
\newcommand{\citeexpid}[2]{({#1}.{#2}.)}

\newcommand{\Children}[1]{\mathit{ch}(#1)}

\newcommand{\setOfBoxes}{\beta}

\newcommand{\intervalArithmOperator}{\mathcal{I\!A}}

\newcommand*\interior[1]{#1^{\mathsf{o}}}

\newcommand{\Fv}{\ensuremath{F_{|\nu}}}

\newcommand{\pn}{\projn{p}}
\newcommand{\pnk}{\projnk{p}}

\newcommand{\projn}[1]{\ensuremath{{#1}_{[1,n]}}}
\newcommand{\projnk}[1]{\ensuremath{{#1}_{[n+1,n+k]}}}

\newcommand{\mysubsetneq}{\ensuremath{\subsetneqq}}

\newcommand{\Cone}{\ensuremath{C^1}-}

\newcommand{\e}{\mathrm{e}}
\newcommand{\dist}{\mathrm{d}}

\begin{document}

\author{Enrico Lipparini\inst{1,2}\orcidlink{0009-0009-0428-4403}\and
	Stefan Ratschan\inst{3} \orcidlink{0000-0003-1710-1513} 
}

\institute{
	University of Genoa, Italy 	\and
	University of Cagliari, Italy \and
	Institute of Computer Science of the Czech Academy of Sciences
}

\setlength{\marginparwidth}{4cm}
\reversemarginpar

\maketitle

\begin{abstract}

\begin{center}
\end{center}
	\noindent
	For typical first-order logical theories, satisfying assignments have a straightforward finite representation that can directly serve as a certificate that a given assignment satisfies the given formula. For non-linear real arithmetic augmented with trigonometric and exponential functions (\nta), however, there is no known direct representation of satisfying assignments that allows for a simple independent check of whether the represented numbers exist and satisfy the given formula. Hence, in this paper, we introduce a different form of satisfiability certificate for \nta, and formulate the satisfiability problem as the problem of searching for such a certificate.
	This does not only ease the independent verification of satisfiability, but also allows the design of new algorithms that show satisfiability by systematically searching for such certificates.          Computational experiments document that the resulting algorithms are able to prove satisfiability of a substantially higher number of benchmark problems than existing methods.         
We also characterize the formulas  whose satisfiability can be demonstrated by such a certificate, by providing lower and upper bounds in terms of relevant well-known classes.
Finally we show the existence of a procedure for checking the satisfiability of \nta-formulas that terminates for formulas that satisfy certain robustness assumptions.        
\end{abstract}


\section{Introduction}
SAT modulo theories (SMT) is the problem of checking whether a given 
first-order formula with both propositional and theory variables is satisfiable in a specific first-order theory. In this paper, we consider the quantifier-free fragment of \smtnta, non-linear real arithmetic  augmented with trigonometric and exponential transcendental functions. 
 This problem is particularly important in the  verification of hybrid systems and in theorem proving.  Unfortunately, \nta is a very challenging theory. 
 Indeed, it is undecidable \cite{richardson68}, and, moreover, there is no known
way of representing satisfying assignments using a finite string of bits (i.e., no finite representation of satisfying assignments)
that could serve as a direct certificate of satisfiability. This does not only make it difficult for an SMT-solver to prove satisfiability, but also raises the question of how to verify the result given by an SMT-solver. 

 In this paper, we

 \begin{itemize}
 \item  introduce the notion of a satisfiability certificate for \nta that allows independent entities to verify the satisfiability of a given input formula without having to re-do a full check of its satisfiability, we 
 \item introduce a method for computing such satisfiability certificates, we
 \item describe computational experiments that analyze the efficiency of several variants of the method, and we
 \item provide a theoretical characterization of the class of problems  for which such a certificate can be successfully computed despite the mentioned undecidability restrictions.
 \end{itemize}

Based on the introduced certificate, the check of satisfiability of \smtnta formulas is both easier in terms of computational effort and effort needed to implement the checker and to ensure its correctness. The certificate is based on the notion of topological degree~\cite{Fonseca:95,Aberth:94,Franek:12b}, generalizing the idea that a sign change of a continuous function $f$ implies satisfiability of $f=0$. The basic tool for checking correctness of the certificate is interval arithmetic~\cite{Rump:10,Neumaier:90,Moore:09}.

 The idea to verify satisfiability of \smtnta in such a way, is not new~\cite{ATVApaper}. However, the formulation as the problem of searching for a certificate is. In addition to the possibility of independent verification, such a formulation makes the corresponding search problem explicit. This allows us to introduce new, efficient search heuristics that guide the algorithm toward finding a certificate and prevent the procedure from getting stuck in computation that later turns out to not to lead to success.


 The experimental results are based on our implementation in the tool~\ugotNL~\cite{ATVApaper}. We compare different heuristic configurations over a wide variety of \nta benchmarks. The benchmarks also demonstrate 
that this new version of \ugotNL outperforms the previous version, making it---to the best of our knowledge---the most effective solver for proving satisfiability of \nta problems.

It is possible to integrate the resulting method into a conflict-driven clause learning (CDCL) type SMT solver~\cite{ATVApaper}. However, in order to keep the focus of the paper on the concern of certificate search, we ignore this possibility, here.

\myparagraph{Content.} The paper is organized as follows: In Section~\ref{sec:preliminaries} we provide the necessary background. In Section~\ref{sec:goal} we give the formal definitions of \emph{certifying SMT solver} and of \emph{satisfiability certificate} in \smtnta. In Section~\ref{sec:method} we outline our method for searching for a certificate, and in Section~\ref{sec:certificate-search} we illustrate   the heuristics that we introduce in detail. In Section~\ref{sec:experiments} we experimentally evaluate our method. In Section~\ref{sec:th_considerations} we provide some theoretical results that characterize
the class of problems that are solvable through the method.
In Section~\ref{sec:relatedwork} we discuss related work. Finally, in Section~\ref{sec:conclusions}, we draw some conclusions.

{\myparagraph{Pre-print version. }
	This is a pre-print version of the paper published in the Journal of Automated Reasoning~\cite{LippariniRatschanJAR}, itself an extended version of a paper that appeared in the proceedings of the NASA Formal Methods Symposium 2023~\cite{Lipparini:23}. 
	The content of this work has also been included in the doctoral thesis of one of the authors~\cite{LippariniThesis}.


\section{Preliminaries}
\label{sec:preliminaries}
We work in the context of \emph{Satisfiability Modulo Theories} (SMT). Our theory of interest is the quantifier-free theory of non-linear real arithmetic augmented with trigonometric and exponential transcendental functions, \smtnta. We assume that the reader is familiar with  standard SMT terminology~\cite{SMTBackground}.

\emph{Notation.} We denote \smtnta-formulas by $\phi, \psi$, clauses by $C_1, C_2$, literals by $l_1, l_2$, real-valued variables by $x_1, x_2,\dots$,
 constants by $a, b$, intervals of real values by $I = [a,b]$, boxes by  $B=I_1 \times \cdots \times I_n$, 
 logical terms with addition, multiplication and transcendental function symbols by $f, g$, and multivariate real functions with $F, G, H$. For any formula $\phi$, we denote by $\Vars{\phi}$ the set of its real-valued variables. When there is no risk of ambiguity we write $f, g$ to also denote  the real-valued functions corresponding to the standard interpretation of the respective terms. We assume that formulas are in Conjunctive Normal Form (CNF) and that their atoms are in the form $f \bowtie 0$, with $\bowtie\;\in \{=, \leq, <\}$. We remove the negation symbol by rewriting every occurrence of $\neg (f = 0)$ as $(f<0 \lor 0< f)$ and distributing $\neg$ over inequalities.

\myparagraph{Points and boxes.} Since we have an order on the real-valued variables $x_1,x_2,\dots$, for any set of variables $V\subseteq \{x_1,x_2,\dots\}$ we can view an assignment $p: V\rightarrow \mathbb{R}$ equivalently as the $|V|$-dimensional point $p\in \mathbb{R}^{|V|}$, and an \textit{interval assignment} $\beta : V \to \{[a,b] : a,b\in \R\}$ equivalently as the $|V|$-dimensional box $B \subseteq \R^{|V|}$. By abuse of notation, we will use both representations interchangeably, using the type $\mathcal{R}^{V}$ both for assignments in $V\rightarrow \mathbb{R}$ and points in $\mathbb{R}^{|V|}$, and the type $\mathcal{B}^{V}$ both for interval assignments in $V\rightarrow \{[a,b] : a,b\in \R\}$ and corresponding boxes. This
will allow us to apply mathematical notions usually defined on points or boxes to such assignments, as well. Given a point  $p\in \mathcal{R}^{V}$, and a subset $V'\subseteq V$, we denote by $\mathit{proj}_{V'}(p)\in \mathcal{R}^{V'}$ the projection of $p$ to the variables in $V'$, that is, for all $v\in V'$, $\mathit{proj}_{V'}(p)(v):= p(v)$. Finally, %
we will use variable assigments $\nu: V\rightarrow \mathbb{Q}$ as substitutions, denoting by $\nu(\phi)$ the result of replacing every variable $v\in V\cap \Vars{\phi}$ in $\phi$ by $\nu(v)$.


\myparagraph{Systems of equations and inequalities.} We say that a formula $\phi$ 
that contains only conjunctions of atoms in the form $f=0$ and $g\leq 0$  is a \emph{system of equations and inequalities}. If $\phi$ contains only equations (inequalities) then we say it is a \emph{system of equations} (\emph{inequalities}). A system of equations ${f_1 = 0 \land \cdots \land f_n=0}$, where the $f_1,\cdots,f_n$ are terms in the variables $x_1, \cdots, x_m$, can be seen in an equivalent way as the equation $F=0$, where $F$ is the real-valued function ${F := f_1 \times \cdots \times f_n : \R^m \to \R^n}$ and $0$ is a compact way to denote the point $(0,\cdots,0)\in \R^n$. 
Analogously, we can see a system of inequalities ${g_1 \leq 0 \land \cdots \land g_k \leq 0}$ as the inequality $G\leq0$,  where $G$ is the real-valued function ${G := g_1 \times \cdots \times g_k : \R^m \to \R^k}$ and $\leq$ is defined element-wise. We will write $eq(\phi)$ for the function $F$ defined by the equations in the formula $\phi$ and ${ineq(\phi)}$ for the function $G$ defined by the inequalities in~$\phi$. We say that a system of equations and inequalities is \emph{bounded} if every free variable appearing in the system has an associated bound consisting of a closed interval with rational endpoints. The handling of strict inequalities would be an easy, but technical extension of our method, which we avoid to stream-line the presentation.

\myparagraph{Dulmage–Mendelsohn decomposition.} Given a system of equations~$\phi$, it is possible to construct an associated bipartite graph $\mathcal{G}_\phi$ that
represents important structural properties of the system. This graph
has one vertex per equation, one vertex
per variable, and an edge between a variable $x_i$ and an equation $f_j=0$ iff $x_i$ appears in $f$. The Dulmage–Mendelsohn decomposition~\cite{dulmage_mendelsohn_1958,dulmage_mendelsohn_system_of_equations} is a canonical decomposition from the field of matching theory that partitions the system into three sub-systems: an over-constrained one (more equalities than variables), an under-constrained one (less equalities than variables), and a well-constrained one (as many equalities as variables, and contains no over-constrained subsystem, i.e. it satisfies the Hall property~\cite{Hall:35}).

\begin{example}
	Let $\phi:= x-\tan(y)=0 \land z^2=0 \land w=0 \land \sin(w)=0$. Through the Dulmage–Mendelsohn decomposition we obtain an under-constrained sub-system $x-\tan(y)=0$ (two variables, one equation), a well-constrained sub-system $z^2=0$ (one variable, one equation), and an over-constrained sub-system $w=0 \land \sin(w)=0$ (one variable, two equations).
\end{example}

\myparagraph{Topological degree.} The notion of the degree of a continuous function (also called the topological degree) comes from differential topology~\cite{Fonseca:95,BrouwerDegreeDincaMawhin}. 
For a continuous function $F: B\subseteq\mathbb{R}^n\rightarrow \mathbb{R}^n$ and a point $p\in \mathbb{R}^n$ such that $p\not\in F(\partial B)$ (where $\partial B$ is the topological boundary of $B$), the degree $\deg(F, B, p)$
is a computable~\cite{Aberth:94,Franek:12b} integer
that  satisfies several interesting properties. The one that plays a pivotal role in our work is the \emph{topological degree test}, which states that: if ${\deg(F, B, 0) \neq 0}$, then the equation $F = 0$ has a solution in $B$.


In the one-dimensional case (i.e., $n=1$ and $B$ being an interval $[a,b]$ with $f(a)\neq p$ and $f(b)\neq p$), the topological degree test is analogous to the corollary of the Intermediate Value Theorem commonly known as Bolzano's Theorem since
 $\deg(F, B, 0)= \frac{\mathrm{sgn}(F(b))-\mathrm{sgn}(F(a))}{2}$. 
 So, for $F(x)=x$, ${\deg(F, [-1,1], 0)=\frac{1-(-1)}{2}=1}$, 
 while for $F(x)=x^2$, $\deg(F, [-1,1], 0)=0$. 

The topological degree generalizes this to functions with $n>1$.
In particular, if $p$ is regular (i.e., for all $y\in B \cap F^{-1}(p)$, $\det F'(y)\neq 0$) then the topological degree of $F$ at $p$ can be defined as
 $\deg(F,B, p) \defas  \sum\limits_{y\in B \cap F^{-1}(p)} \mathrm{sgn}\det F'(y)$. 
This definition can be further extended to non-regular values $p$ in a unique way by continuity of $\deg(F,\Omega, p)$ as a function in $p$, or alternatively, be defined axiomatically (see~\cite{topDegBook} for more details).

While
$\deg(F, B, 0)\neq 0$ implies that $F$ has a root in $B$,
the converse is not true, and the existence of a root does not imply nonzero degree in general. Still, 
if a box contains one isolated zero with non-singular Jacobian matrix, then the topological degree is non-zero~\cite{Fonseca:95}. For alternatives to the topological degree test see our discussion of related work.

\myparagraph{Interval Arithmetic.}
The basic algorithmic tool that underlies our approach is floating point interval arithmetic ($\intervalArithmOperator$)
~\cite{Rump:10,Neumaier:90,Moore:09}
which, given a box~$B$ and an $\nta$-term representing a  function $H$, is able to compute an interval $\intervalArithmOperator_H(B)$ that over-approximates the range $\{ H(x) \mid x\in B\}$ of $H$ over $B$. Since this is based on floating point arithmetic, the time needed for computing $\intervalArithmOperator_H(B)$ does not grow with the size of the involved numbers. Moreover conservative rounding guarantees correctness under the presence of round-off errors.
In the paper, we will use interval arithmetic within topological degree computation~\cite{Franek:12b}, and as a tool to prove the validity of inequalities on boxes.

\myparagraph{Robustness.} We say that a formula $\phi$ is robust if there exists some $\epsilon>0$ such that either every formula that is the result of an $\epsilon$-small perturbation of $\phi$ 
(i.e. a formula whose \emph{distance} from $\phi$ is less than $\epsilon$
) 
is satisfiable, or either every such formula is unsatisfiable. For example, the satisfiable formula $x^2=0$ is not robust, since for every $\epsilon>0$, the perturbed formula $x^2+\epsilon/2=0$ is unsatisfiable. In contrast to that, the satisfiable formula $x^3=0$ is robust.  If $\phi$ is both robust and (un)satisfiable, we say that it is robustly (un)sat. Hence, $x^3=0$ is robustly sat.

For the first part of the paper, an intuitive understanding of the notion of $\epsilon$-small perturbation suffices. In Section~\ref{subsec:BackgroundRobustness}, we will make this more precise for a more formal analysis of our approach. For more details, we refer the reader to the literature~\cite{Franek:12}.
\\
\indent \emph{Relation between robustness and system of equations}: An over-constrained system of equations is never robustly sat~\cite[Lemma 5]{Franek:12}. It easily follows that a system of equations that contains an over-constrained sub-system (in the sense of the Dulmage–Mendelsohn decomposition) is never robustly sat as well.
\\
\indent \emph{Relation between robustness and topological degree}: Even in the case of an isolated zero, the test for non-zero topological degree can fail if the system is non-robust. For example, the function $F(x) \equiv x^2$ has topological degree $0$ in the interval $[-1,1]$%
, although the equality $x^2=0$ has an isolated zero in this interval.
It can be shown that the topological degree test is able to prove satisfiability in all robust cases for a natural formalization of the notion of robustness~\cite{Franek:12}. We will not provide such a formalization, here, but use robustness as an intuitive measure for the potential success when searching for a certificate.

\myparagraph{Logic-To-Optimization.}  
While symbolic methods usually struggle dealing with \nta, numerical methods, albeit inexact, can handle transcendental functions efficiently. For this reason, an SMT solver can benefit from leveraging numerical techniques. In the Logic-To-Optimization approach~\cite{ATVApaper,xsat,SETTA}, 
an \smtnta-formula $\phi$ in $m$ variables 	is translated into a real-valued non-negative function $\LtoO(\phi) \equiv H: \R^{m} \mapsto \R^{\ge 0}$ 
such that---up to a simple translation between Boolean and real values for Boolean variables---each model of $\phi$ is a zero of $H$ (but not vice-versa). When solving a satisfiability problem, one can try to first minimize this function through numerical methods,  then use the obtained numerical (approximate) solution to prove, through exact 
methods, that the logical formula has indeed a model.
 
While for the complete definition of the $\LtoO$ operator we refer to~\cite[Section 3]{ATVApaper}, 
we now provide a simple example to give an intuition on how the operator works. 
Given a formula of the form $F=0$, we have that $\LtoO(F=0) \equiv  F^2$, i.e.,  
for every $x$ in the domain of $F$,  $(\LtoO (F))(x) = F(x)^2$. 
For conjunctions, $\LtoO(F_1 \land F_2) \equiv \LtoO(F_1) + \LtoO(F_2)$.
For disjunctions,  $\LtoO(F_1 \lor F_2) \equiv \LtoO(F_1) * \LtoO(F_2)$

Now, consider for example $F_1, F_2:\mathbb{R}\to \mathbb{R}$ defined by $F_1(x) \equiv x^2-2x$ and $F_2 \equiv \sin(x)$. The formula $F_1=0$ has exactly 2 solutions, $\{0, 2\}$, which are exactly the zeros (hence the global minima) of the non-negative function $F_1^2: \mathbb{R}\to \mathbb{R}$, while the formula $F_2 = 0$ has infinitely many solutions $\{k\pi \ |\  k\in \mathbb{Z} \}$, which are exactly the zeros (and global minima) of $F_2^2: \mathbb{R}\to \mathbb{R}$. Then, in order to find the solutions of $\phi \equiv (F_1 = 0 \land F_2 = 0)$ (which, in this case, consist of the singleton $\{0\}$), we can search for the zeros of $\LtoO(\phi)$. 


\section{Goal}
\label{sec:goal}
Consider an SMT solver that takes as input some formula $\phi$ and as output an element of $\{ \Sat,\Unknown, \Unsat\}$.  How can we gain trust in the correctness of the result of such an SMT solver? One approach would be to ensure that the algorithm itself is correct. Another option is to provide a second algorithm whose output we compare with the original one. Both approaches are, however, very costly, and moreover, the latter approach still may be quite unreliable.

Instead, roughly following McConnell et al.~\cite{McConnell:11} (see also Figure~\ref{fig:certifying}), we require our solver to return---in addition to its result---some information that makes an independent check of this result easy:

\begin{definition}
  \label{def:certifying}
  An SMT solver is \emph{certifying} iff there is a property $W$ such that for every 
   input formula $\phi$, in addition to an element $r\in \{ \Sat, \Unknown, \Unsat \}$, the solver returns an object~$w$ (a \emph{certificate}) such that
	\begin{itemize}
        \item $(\phi, r, w)$ satisfies the property $W$, that is $W(\phi,r,w)$,
        \item     $W(\phi, \Sat, w)$ implies that $\phi$ is satisfiable, 
        \item  $W(\phi, \Unsat, w)$ implies that $\phi$ is unsatisfiable, and
          
		\item   there is an algorithm (a \emph{certificate checker}) that
		\begin{itemize}
			\item takes as input a triple $(\phi, r, w)$ and returns $\top$ iff $W(\phi, r, w)$, and that
			\item is simpler than the SMT solver itself.
		\end{itemize}
		
	\end{itemize}
	
      \end{definition}

\begin{figure}[tbh]
\centering
\includegraphics[clip,trim={0cm 3cm 0cm 0cm},width=8cm]{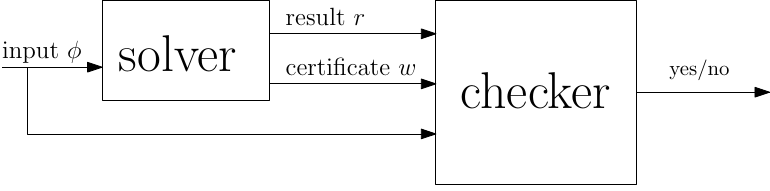}
	\caption{Certifying SMT Solver}
	\label{fig:certifying}
\end{figure}

So, for a given formula $\phi$,  one can ensure correctness of the result $(r, w)$ of a certifying SMT solver by using a certificate checker to check the property $W(\phi, r, w)$. Since the certificate checker is simpler than the SMT solver itself, the correctness check is simpler than the computation of the result itself. 

The definition leaves it open, what precisely is meant by  ``simpler''. In general, it could either refer to the run-time of the checker, or to the effort needed for implementing the certificate checker and ensuring its correctness. The former approach is taken in computational complexity theory, the latter in contexts where correctness is the main concern~\cite{McConnell:11}. Indeed, we will later see that our approach succeeds in satisfying both requirements, although we will not use complexity-theoretic measures of run-time, but will measure run-time experimentally. 

The use of such certificates is ongoing research in the unsatisfiable case~\cite{Barbosa:22}. In the satisfiable case, for most theories, one can simply use satisfying assignments (i.e., witnesses) as certificates. Here the property $W$ simply is the property that the given assignment satisfies the formula, which can be checked easily.

For \smtnta, however, the situation is different. In this case, satisfying assignments may involve numbers that are neither rational nor real algebraic. Indirect descriptions of satisfying assignments using formulations such as
  \begin{itemize}
  \item ``the number that is the solution of the equation $\sin x=1$'', or
  \item ``the number corresponding to the infinite sequence of digits that the Turing machine~$T$ writes onto its output tape''  
  \end{itemize}
  also  cannot be used as a basis for representing certificates. In the first case, the problem is that it is not obvious how to check whether the represented number actually exists. In the second case, it is not obvious how to check whether the represented number actually is a solution.

  In general, adding such information to $\nta$-formulas does not reduce the undecidable satisfiability problem to any known decidable class that would enable a certificate checker in the sense of Definition~\ref{def:certifying}.   Hence one needs to use certificates of a different form. For this, we introduce the following definition:
\begin{definition}
  \label{def:certificate}
Let $\phi$ be a formula in \nta. A \emph{(satisfiability) certificate} for $\phi$ is a triple $(\sigma, \nu, \setOfBoxes)$ such that $W(\phi, \Sat, (\sigma, \nu, \setOfBoxes))$ iff
	\begin{itemize}
	\item $\sigma$ is a function selecting a literal from every clause of $\phi$
		\item $\nu$ is a variable assignment in $\mathcal{R}^{V}$ assigning floating point numbers to a subset $V\subseteq\Vars{\sigma(\phi)}$ (where $\sigma(\phi)$ is a compact way of writing $\bigwedge_{ C \in \phi} \sigma(C)$), s.t. $\sigma(\phi)$
		 contains as many equations as real-valued variables not in $V$.
		\item $\setOfBoxes$ is a finite set of interval assignments in $\mathcal{B}^{\Vars{\phi}\setminus V}$
		such that their set-theoretic union as boxes
		is again a box $B_\beta$ and, 
		for the system of equations  $F:= eq(\nu(\sigma(\phi)))$ 
		and the system of inequalities $G:= ineq(\nu(\sigma(\phi)))$, it holds that:
	
		\begin{itemize}
			\item $0\not\in F(\partial B_\beta)$,
			\item $\deg(F, B_\beta, 0)\neq 0$, and
			\item for every $B\in\setOfBoxes$, $\intervalArithmOperator_G(B)\leq 0$.
		\end{itemize}
	\end{itemize}

\end{definition}

\ 

\begin{example}
\label{ex:certificate}
Consider the formula 
\begin{alignat*}{2}
	& \qquad \qquad \qquad \qquad \qquad \phi := C_1 \land C_2 \land C_3 \land C_4  \\
	& C_1 \ \equiv \ \cos(y) = 0 \ \lor \ \sin(y) = \e^x
	&& C_3 \ \equiv \ x-y \leq \cos(z) \\ 
	& C_2 \ \equiv \  \sin(y)=0 \ \lor \  \cos(y) = \sin(8x^2-z) 
	&& C_4 \ \equiv \ x+y \geq \sin(z) 
      \end{alignat*}      
The following $(\sigma, \nu, \beta)$ is a certificate:
\begin{itemize}
	\item $\sigma := \{ C_1 \mapsto \sin(y) = \e^x\ ;\  C_2 \mapsto \cos(y) = \sin(8x^2-z)\ ; \\ C_3 \mapsto C_3\ ;\  C_4 \mapsto C_4\}$ 
	\item $\nu := \{ z \mapsto 0.2 \}$
	\item $\setOfBoxes:=\{B\}$, where $B := \{x \mapsto [-0.1,0.05]\ ;\ y \mapsto [1.4, 1.9] \} $
        \end{itemize}
      \end{example}

      As can be seen in Figure~\ref{fig:example}, the solution sets of $C_1$ and $C_2$ cross at a unique point in $B$, which reflects the fact that the degree of 
      the function $(x,y)\rightarrow (\sin(y)-\e^x,\cos(y)-\sin(8x^2-0.2))$ is non-zero. Moreover, the inequalities $C_3$ and $C_4$ hold on all elements of the box.
      \begin{figure}[tb]
\centering
        \includegraphics[width=8cm]{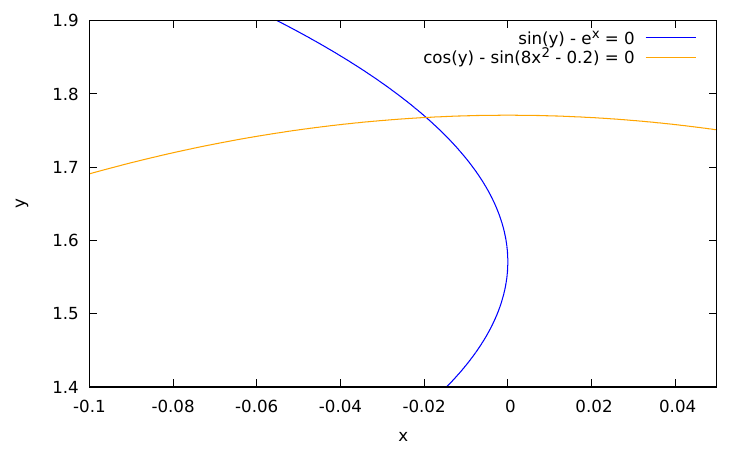}       
        \caption{Solution sets of equalities for the certificate of Example \ref{ex:certificate}}
        \label{fig:example}
      \end{figure}
\ 

Due to the properties of the topological degree and of interval arithmetic discussed in the preliminaries, we have:
\begin{property}
	$W(\phi, \Sat, (\sigma, \nu, \setOfBoxes))$ implies that $\phi$ is satisfiable.
\end{property}

Such a satisfiability certificate  can only serve as a certificate for an SMT solver if a certificate checker---as required by Definition~\ref{def:certifying}---exists. This certificate checker needs to be able to check the conditions of Definition~\ref{def:certificate}. The topological degree can indeed be computed algorithmically~\cite{Aberth:94,Franek:12b}. The condition $0\not\in F(\partial B_\beta)$ is necessary for the topological degree to be well defined. Due to this, such algorithms~\cite{Aberth:94,Franek:12b} also check this condition, and no separate check is necessary. Finally, the condition $\intervalArithmOperator_G(B)\leq 0$ clearly is algorithmic.

Note that this definition could be applied to a broader class of formulas than \nta, 
since the algorithms used to compute the topological degree and $\intervalArithmOperator$ can work with any  \emph{interval-computable} function (i.e. functions for which it is possible to compute arbitrarily precise images for every interval domain~\cite{Franek:12}). 
However, in practice, the tools that implement these algorithms do not go beyond the \nta case.
For this reason, given that our approach is application-oriented, we will keep our focus only on \nta.

In this paper, we will show that in addition to the discussed benefits for correctness, formulating satisfiability checking as the problem of search for such certificates also is beneficial for efficiency of the SMT solver itself. Since we will concentrate on satisfiability, we will simply ignore the case when an SMT solver returns $\Unsat$, so the reader can simply assume that an SMT solver such as the one from Figure~\ref{fig:certifying} only returns an element from the set $\{\Sat, \Unknown\}$.

Note that the variable assignment $\nu$ can also be viewed as a system of equalities of the form $\bigwedge_{v \in V} v=\nu(v)$. In general, one could allow a system of equalities of a more general form, for example, a system of linear equalities, arriving at the function $F$ and $G$ by eliminating some of the involved variables by Gaussian elimination. One could even extend the functions $F$ and $G$ by the left-hand sides of those additional equalities. However, our computational experiments will demonstrate that this is not beneficial, in general, which justifies the more specific form of Definition~\ref{def:certificate}.



\section{Method}
\label{sec:method}
Our goal is to find a triple $(\sigma, \nu, \setOfBoxes)$ that is a certificate of satisfiability for a given formula $\phi$. So we have a search problem. In order to make this search as efficient as possible, we want to guide the search toward a triple that indeed turns out to be a certificate, and for which the corresponding conditions are computationally easy to check.

Intuitively, we view the search for a certificate as a hierarchy of nested search problems, where the levels of this hierarchy correspond to the individual components of certificates. We formalize this using a search tree whose nodes on the $i$-th level are labeled with $i$-tuples containing the first $i$ elements of the tuple searched for, starting with the root note that is labeled with the empty tuple $()$. The tree will be spanned by a function $ch$ that assigns to 
each node $(c_1,\dots, c_i)$ of the tree a sequence $\langle x_1,\dots,x_n\rangle$ of possible choices   for the next tuple component. Hence the children of $(c_1,\dots, c_i)$ in the tree are $(c_1,\dots, c_i, x_1),\dots,(c_1,\dots, c_i, x_n)$. We will do depth-first search in the resulting tree, searching for a leaf labeled by a certificate of satisfiability for the input formula $\phi$.

 Based on the observation that on each level of the tree one has the first $i$ components of the tuple available for determining a good sequence of choices, we will add additional information as the first tuple component in the form of a variable assignment~$p$ that satisfies the formula
 $\phi$ approximately. Hence we search for a four-tuple~$(p, \sigma, \nu, \setOfBoxes)$.

It is easy to see that it would be possible to generalize such a search tree to a more fine-grained one, where the individual levels are formed by parts of the choices described above, and where the order of those levels can be arbitrary. 
For example, it would be possible to first choose an interval for a variable (i.e., part of the box $\beta$), then select a literal from a certain clause (i.e., part of the selection function $\sigma$), and so on. However, in this paper, we keep these levels separated, as discussed above, in order to achieve a clear separation of concerns when exploring design choices at the individual levels.

\section{Certificate Search}
\label{sec:certificate-search}

In this section, we will discuss possibilities for search strategies by defining for every search tree node labeled with tuple  $\tau$, the ordered sequence $\Children{\tau}$ of choices for the next tuple element. Our framework allows for many more possibilities from which we choose strategies that both demonstrate the applicability of the framework to different search strategies, and allow for efficient search, as will be demonstrated by the computational experiments in Section~\ref{sec:experiments}.

In order to be able to refer to different variants of the search strategy in the description of computational experiments, we will introduce keywords for those variants that we will write in teletype font.

Here we will focus on strategies of the two following basic types: 
\begin{itemize}
\item Filtering: We skip elements from the set of possible choices that cannot result in a certificate or for which the probability of resulting in a certificate is negligible.
\item Ordering: We choose elements from the set of possible choices in a certain order that tries the reflect the probability of resulting in a certificate.
\end{itemize}

\subsection{Points}

The points $\Children{}=\langle p_1,\dots, p_k\rangle$ determining the first level of the search tree are generated by an optimization problem defined on the formula $\phi$
following the Logic-To-Optimization approach~\cite{ATVApaper}. Here we translate the satisfiability problem into a numerical minimization problem, mapping the logic formula $\phi$ into the non-negative real-valued function $\LtoO(\phi) \equiv H: \mathbb{R}^n\rightarrow\mathbb{R}_{\geq 0}$
(called the \textit{objective function}) such that for every satisfying assignment, this objective function is zero, and for assignments that do not satisfy the formula, the objective function is
typically (but not always) non-zero. Then we find local minima of $H$ through an unconstrained optimization algorithm such as basin hopping~\cite{basinhoppin}, a two-phases Monte Carlo Markov Chain method that alternates local minimization with random jumps. In our implementation, we compute $k=100$ local minima, and process them in the order of their value.

\subsection{Literals}

Given a point $p$, we now discuss how choose literal selector functions $\Children{p}=\langle \sigma_1,\dots, \sigma_k\rangle$.
For filtering the set of literal selector functions, we will restrict ourselves, 
for each clause $C\in\phi$, to the literals $l$ for which the objective function restricted to $l$ and evaluated in the point~$p$ is below a certain threshold. That is, we determine the set of approximately satisfiable literals $$L_C := \{l\in C \mid \LtoO(l)(p)\leq \epsilon  \}.$$ 
Our literal selector functions will then correspond to the set of all approximately satisfiable combinations $$\{\sigma \mid \text{for all } C\in\phi, \sigma(C)\in L_C \},$$ that is, each $\sigma$ selects exactly one approximately satisfiable literal from each clause. In order to maximize the chances of choosing a better literal combination, 
we can sort the elements of $L_C$
according to the value of the respective objective functions and then choose literal combinations using the corresponding lexicographic order (we will refer to this heuristic as \sortWrtCost).

While the point $p$ is usually a good candidate in terms of \emph{distance from a zero}, it can sometimes lead to an inconsistent problem:
\begin{example}Consider the formula	
	\begin{alignat*}{2}
		& \qquad \qquad \qquad \qquad \qquad \qquad \phi := C_1 \land C_2  \\
		& C_1 \equiv (x+y=0) \lor (x=\e^{10^6*y})
		&& C_2 \equiv (x+y \geq \epsilon_1) \lor (x=\tan(y+\epsilon_1)) 
	\end{alignat*}  	
	The numerical optimizer will be tempted to return first some point $p_1$ such as $\{x\mapsto 1; y\mapsto -1\}$, that \emph{almost} satisfies $(x+y=0) \land (x+y \geq \epsilon_1)$, instead of a harder approximate solution involving transcendental functions and heavy approximations, such as $(x=\e^{10^6*y}) \land (x=\tan(y+\epsilon_1))$, that is exactly satisfiable in a point $p_2$ near $(0, -\pi)$.
\end{example}

Such inconsistencies may occur in many combinations of literals. We use a strategy that detects them in situations where for certain clauses $C$, the set $L_C$ contains only one literal $l$. We will call such a literal $l$ a \emph{forced literal}, since, for every literal selector function~$\sigma$, $\sigma(\phi)$ will include $l$. Before starting to tackle every approximately satisfiable literal combination, we first analyze the set of forced literals. We do symbolic simplifications (such as rewriting and Gaussian elimination) to check whether the set has inconsistencies that can be found at a symbolic level (as in the previous example). If the symbolic simplifications detect that the forced literals are inconsistent then we set $\Children{p}$ to the empty sequence $\langle\rangle$ which causes backtracking in depth-first search. We refer to the variant of the algorithm using this check as \checkForcedLiterals.

\ 

\emph{Filtering out over-constrained systems}. Given a literal selector function~$\sigma$, we analyze the structure of the system of equations formed by the equations selected by $\sigma$ through the Dulmage–Mendelsohn decomposition, that uniquely decomposes the system into a well-constrained subsystem, an over-constrained subsystem and an under-constrained subsystem.
We filter out every literal combination having a non-empty over-constrained subsystem, since this leads to a non-robust sub-problem, referring to this heuristic as \filterOverconstr.

\subsection{Instantiations}

\label{subsec:instantiations}
We define the instantiations $\Children{p, \sigma}=\langle \nu_1,\dots,\nu_k\rangle$ based on a sequence of sets of variables $V_1,\dots,V_k$ to instantiate, and define $\nu_i:= \mathit{proj}_{V_i}(p)$. The uninstantiated part of $p$ after projection to a set of variables $V_i$ is  then $\mathit{proj}_{\Vars{\phi}\setminus V_i}(p)$, which we will denote by $p_{\neg V_i}$.

For searching for the variables to instantiate, we use the Dulmage–Mendelsohn decomposition constructed in the previous level of the hierarchy. We do not want to instantiate variables appearing in the well-constrained sub-system, since doing so would make the resulting system after the instantiation over-constrained. Hence the variables to be instantiated should be chosen only from the variables occurring in the under-constrained subsystem. This substantially reduces the number of variable combinations that we can try. 
Denoting the variables satisfying this criterion by $V_{under}$, this restricts $V_i\subseteq V_{under}$, for all $i\in \{1, \dots, k\}$. 
This does not yet guarantee that every chosen variable combination leads to a well-constrained system after the instantiation. For example
, the under-determined system of equations  $x+y = 0 \land z+w=0$ has four variables and two equations, but becomes over-constrained after instantiating either the two variables $x$ and $y$, or the variables $z$ and $w$. So, for each $V_i$, we further check whether the system obtained after the instantiation is well-constrained (we refer to this heuristic as \filterOverconstrV). 

The method described in the previous paragraph only uses information about which equations in the system contain which variables (i.e., it deals only with the \textit{structure} of the system, not with its \textit{content}). Indeed, it ignores the point~$p$. 

To extract more information, we use the following fact: If a zero of a function has non-singular Jacobian matrix, then every box containing this zero and no other zeros has a non-zero topological degree~\cite{Fonseca:95}.
So we compute a floating point approximation of the Jacobian matrix at point $p$ (note that, in general, this matrix is non-square). Our goal is to find a set of variables $V$ to instantiate such that the Jacobian matrix corresponding to the resulting square system at the point $p_{\neg V}$ has full rank. This matrix is the square sub-matrix of the original Jacobian matrix that is the result of removing the instantiated columns.

A straight-forward way of applying the Jacobian criterion is, given random variable instantiations, to filter out instantiations whose corresponding Jacobian matrix is rank-deficient \filterRankDeficient, similarly to what is done in the previous paragraph with the overconstrained filter. 
Note that, as the Jacobian matrix of non-well-constrained system of equations is always rank-deficient,
this filter is stronger than the previous one. However, it may filter out variable instantiations that result in a non-zero degree (e.g., the function $x^3$ has non-zero degree in $[-1,1]$, but its Jacobian matrix at the origin is rank deficient since $f'(0)=0$).

We can further use the information given by the Jacobian matrix not only to filter out bad variable instantiations, but also to maximize the chance of choosing good variable instantiations from the beginning. Indeed, not all variable instantiations will be equally promising, and it makes sense to head for an instantiation such that the resulting square matrix not only has full rank, but---in addition---is far from being rank-deficient (i.e., it is as robust as possible). 
We can do so by modifying Kearfott's method~\cite[Method 2]{Kearfott:98}, which fixes the coordinates most tangential to the orthogonal hyperplane of $F$ in $p$ by  first computing an approximate basis of the null space of the Jacobian matrix in the point, and then choosing the variables corresponding to the coordinates for which the sum of the absolute values in the basis is maximal.
Since we are interested in more than just a single variable choice, we order all the variables w.r.t. to this sum. Then, we extract the  sets of variables $V_1, V_2, \dots$ through a lexicographic combinatorial algorithm. We refer to this heuristic as \KearfottOrdering.

\myparagraph{Adding equations.} 
\label{subsubsec:orth}
An alternative approach for reducing from an underconstrained system of equations to a square one is, instead of instantiating variables, to add equations. 
This approach is justified by the fact that 
each variable instantiation can be seen as a system of equalities (while the vice-versa is not true). 
 
While discussing Kearfott's method, we showed that, given a point $p$, it is better to choose variable instantiations that are the most orthogonal possible to the tangent hyperplane of $F$ in $p$. 
With the equations adding approach we can go further: we can directly choose the linear equations that describe the hyperplane orthogonal to the tangent space of $F$ in $p$. 
These equations can be found through the QR-decomposition of the Jacobian matrix of $F$ in $p$.
We can then add these equations to $F$ in order to obtain a square system of equations. We refer to this heuristic as \orthogonal.

Since the found equations are linear, we can further modify the previous heuristic by applying Gaussian elimination to the linear part of the square system obtained, thus reducing the dimension of the system of equations. We refer to this sub-heuristic as \gaussel.

\subsection{Boxes}
\label{subsec:box}
We construct boxes around $p_{\neg V}$, where $V$ is the set of variables $\nu$ instantiates, that is, $\nu\in\mathcal{R}^{V}$. So we define $\Children{p, \sigma,\nu}:=\langle \beta_1,\dots,\beta_k\rangle$ s.t. for all $i\in\{1,\dots,k\}$, for all $B\in\beta_i$, $B\in \mathcal{B}^{\Vars{\phi}-V}$ and $p_{\neg V}\in\bigcup_{B\in \beta_i} B$.

We use two different methods, \epsInflation and \boxGridding:
\begin{itemize}
	\item Epsilon-inflation~\cite{Mayer:94} is a method to construct incrementally larger boxes around a point. In this case, the $\beta_1,\dots, \beta_k$ will each just contain one single box $B_i$ defined as the box centered at $p_{\neg V}$ having side length $2^i\epsilon$, where, in our setting, $\epsilon=10^{-20}$. We terminate the iteration if either $\intervalArithmOperator_G(B_i)\leq0$ and $\deg(F,B_i,0)\neq0$, in which case we found a certificate, or we reach an iteration limit (in our setting when $2^i\epsilon > 1$).
	\item Box-gridding is a well-known technique from the field of interval arithmetic based on iteratively refining a starting box into smaller sub-boxes. Here we use a specific version, first proposed in \cite{Franek:12} and then implemented with some changes in \cite{ATVApaper}. In the following we roughly outline the idea behind the algorithm, and refer to the other two papers for details. We start with a grid that initially contains a starting box (in our setting, having side length $1$). We then iteratively refine the grid by splitting the starting box into smaller sub-boxes. At each step, for each sub-box~$B$ 
	we first check whether interval arithmetic can prove that the inequalities or the equations are unsatisfiable, and, if so, we remove $B$ from the grid.
	We check also whether $\deg(F,B,0)\neq0$ and interval arithmetic can prove the satisfiability of the inequalities, and, if so, then we  terminate our search, finding a certificate with the singleton $\beta_i=\{ B\}$. 
	In some cases, in order to verify  the satisfiability of the inequalities, we will have to further split the box $B$ into sub-boxes, using the set of resulting sub-boxes instead of the singleton $\{ B \}$. 
	After each step, if there are sub-boxes left in the grid, we continue the refinement process. Otherwise, if the grid is empty, we conclude that there cannot be solutions in the starting box. If a certain limit to the grid size is exceeded, we also stop the box gridding procedure without success.
      \end{itemize}

For both methods, if the method stops without success, we have arrived at the last element of the sequence of choices $\langle \beta_1,\dots,\beta_k\rangle$ without finding a certificate, which results in backtracking of the depth-first search for a certificate.

Both mentioned methods have their advantages, and can be seen as complementary. Epsilon-inflation is quite fast, and performs particularly well if the solution is isolated and is near the center. However, if there are multiple solutions in a box, the topological degree test can potentially fail to detect them\footnote{For example, for $f(x)=x^2-1$, $deg(f, [-10,10],0)=0$, while $deg(f,[-10,0],0)=-1$, and $deg(f,[0,10],0)=1$.}, and if the solution is far from the center then we need a bigger box to encompass it, which is less likely  to be successful than a smaller box, as we require the inequalities to hold everywhere in the box, and, moreover, the chance of encompassing other solutions (thus incurring in the previous problem) grows. 

The box-gridding procedure, on the other side, can be quite slow, as in the worst case the number of sub-boxes explodes exponentially. However,
grid refinement leads to a very accurate box search, which allows us to avoid the issues faced with epsilon inflation (i.e. multiple solutions, or a solution far from the center). Moreover, if the problem is robust, we have the theoretical guarantee that the procedure will eventually converge to a solution~\cite{Franek:12}, although this does not hold in practice due to the introduced stopping criterion.

Indeed, a third approach is to combine the two methods: first use epsilon inflation, that is often able to quickly find a successful box, and, if it fails, then use the more accurate box-gridding procedure.


\section{Computational Experiments}
\label{sec:experiments}

\begin{table}
	\centering
		\begin{tabular}{|c ||c|c|c|c| } 
			\hline
			
			& \multicolumn{3}{c|}{Heuristics} & (id.) \\ 
			\hline 
			
			N. solved & Literals & Instantiations & Boxes &   \\
			
			\hhline{|=||=|=|=|=|}
			
			323 &   &  & \boxGridding & \expid{1}{a} \\ 	\hline
			
			355 &   &  & \epsInflation &  \expid{1}{b} \\ 	\hline
			
			356 &   &  & \makecell{\epsInflation \\ \boxGridding }&  \expid{1}{c} \\ \hhline{|=||=|=|=|=|}
			
			362 & \sortWrtCost & & \epsInflation &\expid{2}{b} \\ \hline
			
			361 & \sortWrtCost & & \makecell{\epsInflation  \\ \boxGridding }& \expid{2}{c} \\
			
			 \hhline{|=||=|=|=|=|}
			 
			 370 & \makecell{ \sortWrtCost \\ \filterOverconstr} & & \epsInflation & \expid{3}{b} \\ \hline
			 
			 367 & \makecell{ \sortWrtCost \\ \filterOverconstr} & & \makecell{\epsInflation \\ \boxGridding }& \expid{3}{c} \\
			 
			 \hhline{|=||=|=|=|=|}
			 
			 406 & \makecell{ \sortWrtCost \\ \filterOverconstr \\ \checkForcedLiterals} & & \epsInflation & \expid{4}{b} \\ \hline
			 
			 410 & \makecell{ \sortWrtCost \\ \filterOverconstr \\ \checkForcedLiterals} & & \makecell{\epsInflation \\ \boxGridding }& \expid{4}{c} \\
			 
			 
			 
			 
			 \hhline{|=||=|=|=|=|}		
			 	 
			 409 & \makecell{ \sortWrtCost \\ \filterOverconstr \\ \checkForcedLiterals} & \KearfottOrdering & \epsInflation & \expid{5}{b} \\ \hline
			 
			 412 & \makecell{ \sortWrtCost \\ \filterOverconstr \\ \checkForcedLiterals} & \KearfottOrdering & \makecell{\epsInflation \\ \boxGridding }& \expid{5}{c} \\
			 
			 \hhline{|=||=|=|=|=|}
			 
			 424 & \makecell{ \sortWrtCost \\ \filterOverconstr \\ \checkForcedLiterals} & \makecell{ \KearfottOrdering \\ \filterOverconstrV}& \epsInflation & \expid{6}{b} \\ \hline
			 
			 426 & \makecell{ \sortWrtCost \\ \filterOverconstr \\ \checkForcedLiterals} & \makecell{ \KearfottOrdering \\ \filterOverconstrV} & \makecell{\epsInflation \\ \boxGridding }& \expid{6}{c} \\
			 
			 \hhline{|=||=|=|=|=|}
			 
			 427 & \makecell{ \sortWrtCost \\ \filterOverconstr \\ \checkForcedLiterals} & \makecell{ \KearfottOrdering \\ \filterOverconstrV \\ \filterRankDeficient}& \epsInflation & \expid{7}{b} \\ \hline
			 
			 426 & \makecell{ \sortWrtCost \\ \filterOverconstr \\ \checkForcedLiterals} & \makecell{ \KearfottOrdering \\ \filterOverconstrV \\ \filterRankDeficient} & \makecell{\epsInflation \\ \boxGridding }& \expid{7}{c} \\
			 
			 
			 
			 
			 
			 \hhline{|=||=|=|=|=|}			
			 441 & \multicolumn{3}{c|}{Virtual best} & \\
			 \hline	
			 
		\end{tabular}
	
	\caption{\label{fig:experiments} Summary of the results for different heuristics configurations. Each row correspond to a configuration. The  first column from the left contains the number of benchmarks solved; the central columns indicate the heuristics used, separated by search level; the last column contains an identifier of the configuration. The last row is for the virtual best of the different configurations.}
\end{table}


\myparagraph{Implementation.} 
We implemented the different heuristics presented in the paper in a prototype tool called \ugotNL (firstly presented in \cite{ATVApaper}). In order to make the results comparable with the ones obtained earlier, in addition to the search method discussed in Section~\ref{sec:certificate-search}, we preserve the following heuristics used by \ugotNL: If the local minimizer cannot find any minimum of $\LtoO(\phi)$ for which for every clause $C\in\phi$, the set of approximately satisfiable literals $L_C$ is non-empty, we restart the procedure on every conjunction resulting from the DNF of $\phi$.
The tool handles strict inequalities of the form $f<0$ directly until the box construction phase, where they are replaced by $f\leq -\varepsilon$  (with $\varepsilon=10^{-20}$).
For computing the topological degree, we use \textsc{TopDeg}\footnote{Available at \url{https://www.cs.cas.cz/~ratschan/topdeg/topdeg.html}.}. For the symbolic simplifications used in \checkForcedLiterals, we use the \textit{simplify} and the \textit{solve-eqs} tactics provided by \textsc{Z3}~\cite{z3}
\footnote{For a description of the two tactics: \url{https://microsoft.github.io/z3guide/docs/strategies/summary}. The version of Z3 used is 4.5.1.0.}. For the computation of the rank used in \filterRankDeficient, we observe that the rank of a matrix is equal to the number of non-zero singular values, hence we consider a matrix far from rank-deficiency iff all its singular values are bigger than some threshold (to account for approximation errors). We use a threshold widely used by algorithms for determining the matrix rank, which is $\sigma_{\max} dim(A) \varepsilon$, where $\sigma_{\max}$ is the largest singular value of $A$, and $\varepsilon$ is the machine epsilon.

\myparagraph{Setup.} We run the experiments\footnote{The results of the experiments are available at \url{https://doi.org/10.5281/zenodo.7774117}} on a cluster of identical machines equipped with 2.6GHz AMD Opteron 6238 processors.	We set a time limit of 1000 seconds, and a memory limit of 2Gb. We considered all \smtnta benchmarks from the dReal distribution~\cite{dreal} and other \smtnta benchmarks coming from the discretization of Bounded Model Checking of hybrid automata~\cite{HARE,HYST}, totaling 1931 benchmarks. All of these benchmarks come with ``unknown'' status. According to experiments performed on other solvers ($\cvctool$, $\drealtool$, $\isattool$, $\mathsat$), among these benchmarks 736 (respectively, 174) are claimed to be unsatisfiable (satisfiable) by at least one solver\footnote{For the results of such experiments, see \cite{ATVApaper}.}. 
We tested our tool with different heuristics configurations (Table \ref{fig:experiments}), and, for each configuration, we checked that our tool never contradicts the other tools. We have arranged the heuristics into 3 columns (Literals, Instantiations, and Boxes) according to the search level they are used in. As the number of possible configurations is quite high, we proceed as follows: We start with the simpler configurations (just one method for finding a box that contains a solution), and then we add heuristics.
The configurations that use the equation adding method are discussed separately at the end of the section.

\begin{figure}[t]
	\includegraphics[scale=0.8]{"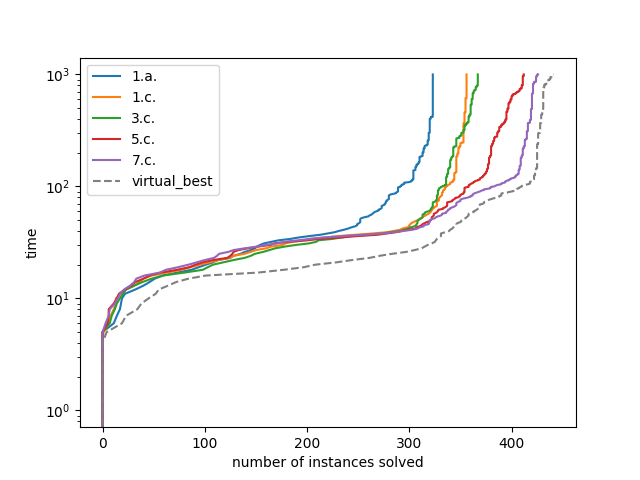"}
	\caption{Survival plots of some of the  configurations presented in Table \ref{fig:experiments}. For each configuration, the plot shows the number of
		instances solved (x axis) within the given time (y axis).}
	\label{fig:cactus}
\end{figure}

\myparagraph{Results.} In the first configurations we tested the 3 possible ways to search for a box. We note that \boxGridding \citeexpid{1}{a} performs  considerably worse than the other two, \epsInflation \citeexpid{1}{b} and \epsInflation+\boxGridding \citeexpid{1}{c}, which produce comparable results. Because of that, and for readability's sake, we did not use \boxGridding alone with other heuristics in the next configurations, but only considered the other two options. We then added heuristics based on the following criteria: first heuristics for the ``Literals'' choice, then heuristics for the ``Instantiations'' choice, and first ordering heuristics (i.e. \sortWrtCost and \KearfottOrdering), then filtering heuristics (all the others). At every new heuristic added, we see that the number of benchmarks solved grows regardless of the ``Boxes'' choice, with the best configuration reaching 427 benchmarks using 7 heuristics. If we consider the virtual best (i.e. run in parallel all the configurations and stop as soon as a certificate is found) we are able to solve 441 benchmarks. This is because in cases such as \epsInflation vs. \epsInflation+\boxGridding, or such as \filterOverconstrV vs. \filterRankDeficient, there is no dominant choice, with each configuration solving benchmarks that the other does not solve and vice-versa. 
The cactus plot in Figure~\ref{fig:cactus}---in which we included only a subset of the configurations in Table~\ref{fig:experiments} for graphical reasons--- 
substantiates the claim that new heuristics improve not only effectiveness, but also performances. 
The plot of the virtual best remarks the complementarity between different configurations.

\begin{figure}[t]
	\includegraphics[scale=0.8]{"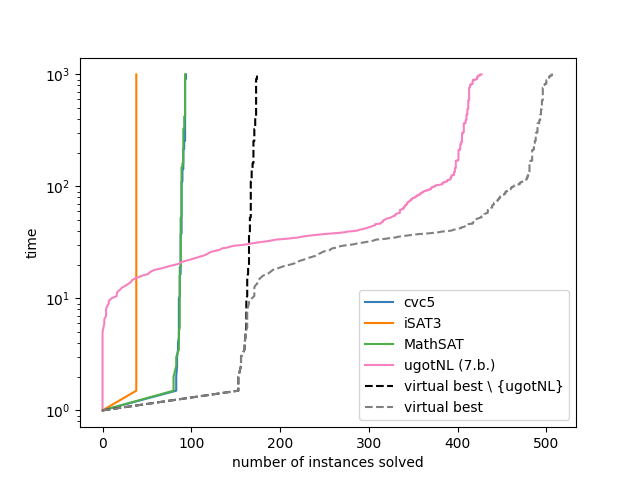"}
	\caption{Survival plots of the best configuration of \ugotNL (\citeexpid{7}{b}) compared with the other state-of-the-art SMT solvers.  
		For each solver, the plot shows the number of
		instances solved (x axis) within the given time (y axis). 
		Note that the plots of \cvctool and \mathsat are very close to each other (this is not surprising, as both tool use the same technique, incremental linearization).
	}
	\label{fig:cactusCompare}
\end{figure}

\myparagraph{Discussion.}  The first configuration \citeexpid{1}{a} essentially uses a method proposed earlier~\cite{ATVApaper} and implemented in a tool called \ugotNLeager (of which the tool presented in this paper is an upgrade). Already in the previous paper, \ugotNLeager outperformed the other solvers able to prove satisfiability in $\smtnta$, solving more than three times the benchmarks than \mathsat\cite{mathsat5}, \cvctool \cite{cvc5}, and \isattool\cite{iSAT3}, and almost as twice as the benchmarks solved by the \emph{lazy} version \msatUgot (where \ugotNL had been integrated \emph{lazily} inside \mathsat). The introduction of new heuristics further improved the performances of our tool, that is now able to solve around 100 benchmarks more.
Moreover, the best configuration of our tool, \citeexpid{7}{b}, was able to prove the satisfiability of 334 benchmarks among the 1021 that had status ``unknown'', i.e. that had not been solved by any other solver. 
Now, considering all the state-of-the-art SMT solvers that are able to prove satisfiability of \nta benchmarks, including ours, 
the virtual best 
 is able to solve 508 benchmarks, compared to the 174 previously solved without the inclusion of our tool. This is depicted in Figure \ref{fig:cactusCompare}, which also shows the complementarity between the different solvers. As can be noted, 
 the plot of \ugotNL is less steep than the plots of other solvers. This is in part due to the eager approach of the tool, that first generates several candidate points, and only then chooses one of these points to narrow down the search, in part due to the tool been a prototype written in Python, compared to highly optimized solvers written in C/C++. 

\myparagraph{Run-time of the certificate checker.} In Section \ref{sec:goal} we claimed that, with our approach, checking a certificate requires less run-time than the certificate search itself. Here we experimentally quantify this amount: for each benchmark solved by the best configuration \citeexpid{7}{b}, we observe the run-time required to check the certificate (which amounts, essentially, to the computation of topological degree and interval arithmetic for the successful box). In terms of median (respectively, mean), checking the certificate requires $0.10\%$ ($1.07\%$) of the run-time used by the solver.

\myparagraph{Variable instantiations vs. equation adding}
In Section \ref{subsec:instantiations} we presented two different approach to reduce to a square system of equations: variable instantiation and equation adding. 
In Table \ref{fig:experimentsOrth} we experimentally compare these two approaches. In order not to overload the table, we compare the new heuristics only against representative configurations: \citeexpid{7}{b} - being the one that solved more benchmarks - and \citeexpid{4}{b} - being the one where only "Literals" heuristics are used. 
We observe that the use of \orthogonal does not seem to pay off particularly well: comparing \citeexpid{4}{b} to \citeexpid{orth}{1}, it increases the number of benchmarks solved just by a small margin, while comparing \citeexpid{7}{b} to \citeexpid{orth}{3}, the number of benchmarks solved decreases. If we consider the configurations that use the sub-heuristic \gaussel\ - \citeexpid{orth}{2} and \citeexpid{orth}{4} - we see that in both cases the performance are even worsened.

\myparagraphB{Discussion.} While from a mathematical perspective the orthogonal method should perform better, as it yields a more robust system and obviates the iteration through all the possible variable instantiations, 
its effectiveness can be severely limited by the well-known numerical instability of algorithms that compute orthogonal matrices. 
Moreover, the systems of equations obtained via equation adding have a higher dimension than the ones obtained via variable instantiations, leading to a more complex problem to solve for the topological degree test and interval arithmetic.
One could hope to lighten this negative effect by using the sub-heuristic \gaussel  
that reduces the dimension to the same obtained via variable instantiation. 
Unfortunately, this procedure generates equations which are more complex then the ones obtained by variable instantiation, 
as it substitutes the exceeding variables by a linear combination of the remaining variables, 
and since our technique for proving satisfiability relies on interval arithmetic 
(which is quite sensitive to syntactic manipulations and rounding bounds propagation)
this can severely impact on the effectiveness of the check.
While we do not rule out that a more engineered implementation of the equation adding approach could yield better results, our experimental results show that the more straight-forward approach of variable instantiation is more effective on the considered benchmarks.

\begin{table}
	
	\centering
		\begin{tabular}{|c ||c|c|c|c| } 
			\hline
			
			& \multicolumn{3}{c|}{Heuristics} & (id.) \\ 
			\hline 
			
			N. solved & Literals & Instantiations & Boxes &   \\

			\hhline{|=||=|=|=|=|}
			
			406 & \makecell{ \sortWrtCost \\ \filterOverconstr \\ \checkForcedLiterals} & & \epsInflation & \expid{4}{b} \\ \hline
			
			409 & \makecell{ \sortWrtCost \\ \filterOverconstr \\ \checkForcedLiterals} & \orthogonal & \epsInflation & \expid{orth}{1} \\ \hline
			
			399 & \makecell{ \sortWrtCost \\ \filterOverconstr \\ \checkForcedLiterals} &  \makecell{\orthogonal  \gaussel } & \epsInflation & \expid{orth}{2} \\ \hline
			
			\hhline{|=||=|=|=|=|}

			427 & \makecell{ \sortWrtCost \\ \filterOverconstr \\ \checkForcedLiterals} & \makecell{ \KearfottOrdering \\ \filterOverconstrV \\ \filterRankDeficient}& \epsInflation & \expid{7}{b} \\ \hline

			 419 & \makecell{ \sortWrtCost \\ \filterOverconstr \\ \checkForcedLiterals} & \makecell{\orthogonal \\ \KearfottOrdering \\ \filterOverconstrV \\ \filterRankDeficient}& \epsInflation & \expid{orth}{3} \\
			
			\hline
			
			 413 & \makecell{ \sortWrtCost \\ \filterOverconstr \\ \checkForcedLiterals} & \makecell{\orthogonal \gaussel \\ \KearfottOrdering \\ \filterOverconstrV \\ \filterRankDeficient}& \epsInflation & \expid{orth}{4} \\ \hline
			
			\hhline{|=||=|=|=|=|}			
			443 & \multicolumn{3}{c|}{Virtual best } & \\
			\hline	
			
		\end{tabular}
	
	\caption{\label{fig:experimentsOrth} Summary of the heuristics configurations using the equation adding approach. Two configurations from from Table \ref{fig:experiments} are included for comparison. The last row is the virtual best considering all the configurations.
	}
\end{table}




\section{Theoretical Characterization}
\label{sec:th_considerations}

Since the problem addressed by this paper is undecidable, the success of any algorithmic approach to solving the problem must necessarily depend on heuristics. Still, in this section, we contribute some results that provide insight into when one can reasonably expect an approach such as the one presented in this paper to succeed. 

Especially, we address a sensitive part of our method---the reduction from an under-constrained system of equations to a well-constrained subsystem (Section \ref{subsec:instantiations}). Indeed, given a system of equations in $m$ variables and $n$ equations ($m>n$), 
in order to obtain a well-constrained system, we need to either instantiate $k:= m-n$ variables, or, alternatively, to add $k$ equations. The contribution of this section is threefold:
\begin{itemize}
	\item We bound the class of problems solvable through the variable instantiation method, both from below and from above.
	\item We prove that the class of problems solvable through the variable instantiation method is a (possibly non-strict) subset of the class of problems solvable through the equation adding method.
	\item We show that for bounded systems of equations and inequalities a certain strategy in the certificate search method presented in Section~\ref{sec:method} will always succeed in determining satisfiability under certain robustness assumptions.%
\end{itemize}

For ease of discussion, first, we will 	consider only systems of equations (i.e., without inequalities). It is however straightforward to see that including inequalities does not change any of the results. 
We will then treat the general case of conjunction and disjunctions of systems of equations and inequalities 	when discussing the last contribution. 

We will start by introducing some notation for the relevant classes of problems. For now, we will introduce these classes only informally, and define them precisely, later. We will denote by $\FRobI$ and $\FRobLEq$ the problem classes, for which the two methods (instantiation and adding equations, respectively) result in a robust system. More precisely, we will denote by  $\FRobI$ 
the class of \Cone functions $F:B\subseteq \mathbb{R}^{n+k}\to \mathbb{R}^{n}$ 
for which there exists a point $p\in \mathbb{R}^{n+k}$ such that the instantiation of $k$ variables to the corresponding $k$ values of $p$ leads to a robust system in $\mathbb{R}^n\rightarrow\mathbb{R}^n$ (we will call such functions \emph{robust under instantiation}), and we will denote by $\FRobLEq$ the class of \Cone functions $F:B\subseteq \mathbb{R}^{n+k}\to \mathbb{R}^{n}$  for which adding $k$ linear equations leads to a robust system in  $\mathbb{R}^{n+k}\rightarrow\mathbb{R}^{n+k}$.

First, we will bound $\FRobI$ from above by the class $\FRob$
of \Cone functions $F:B\subseteq \mathbb{R}^{n+k}\to \mathbb{R}^{n}$ that have 
 a robust solution, and from below by
the class $\FReg$ of \Cone functions 
$F:B\subseteq \mathbb{R}^{n+k}\to \mathbb{R}^{n}$ 
that have a solution that is regular in the sense of topology.

Based on this, we will prove the following:
\begin{theorem}
	\label{thm:FRobIbounds}
	$\FReg \mysubsetneq  \FRobI  \mysubsetneq \FRob $ 
\end{theorem}

As can be seen from the disequalities, the lower and upper bounds are strict, here.


Secondly, we prove that every problem that can be solved via variable instantiation can be solved via adding equation, i.e. we have the following theorem:

\begin{theorem}
\label{thm:instVersusEq}      
	$  \FRobI  \subseteq \FRobLEq $ 
\end{theorem}

Note that the inclusion, in this case, is not strict. Indeed, we conjecture that the equality holds, but will leave the proof to future work.

Finally, we present a variation of our method that is guaranteed to always terminate on problems in $\FRobI$, and that will serve to prove the following theorem:

\begin{theorem}
	\label{thm:quasiquasidecidability}
	There exists a procedure that, given a bounded system of equations and inequalities $F=0 \wedge G \leq 0$,
	\begin{itemize}
        \item always returns the correct answer ``satisfiable'' or ``unsatisfiable'', if it terminates, 
        \item always terminates successfully when $F=0 \wedge G \leq 0$ is robustly satisfiable  and  $F\in \FRobI$,
        \item always terminates successfully when $F=0 \wedge G \leq 0$ is robustly unsatisfiable.
	\end{itemize}
\end{theorem}

This theorem can be seen as an extension of an earlier result~\cite{Franek:12} showing
that the class of bounded systems of equations and inequalities in $m$ variables and $n$ equations, with $n\geq m$ or $n=0$, is quasi-decidable in the sense that there exists a procedure that always terminates on robust instances, and that never returns a wrong answer.
 
Our contribution is to cover the case of under-constrained systems (i.e. when $n < m$), to a certain extent. 
Indeed, it is not possible to simply remove the restriction on the number of equations versus the number of variables~\cite[Theorem 2]{Franek:12}. 
We overcome this by guaranteeing termination in the satisfiable case only for problems for which the system of equations is robust under instantiation (which is a stricter condition than general robustness). 
In this sense, our procedure is not a quasi-decision procedure, 
as it does not terminate for \emph{all} robust instances, 
but it will cover a meaningful sub-class.

\

The section is organized as follows. In Section \ref{subsec:BackgroundRobustness}, we formalize
the problem classes mentioned by the two theorems and provide some further 
definitions and properties that will be useful in the following subsections.
In Section \ref{subsec:GuaranteesRegularity},  we will prove the lower bound $\FReg \subseteq \FRobI$ (Lemma \ref{thm:FRegsubsetFRob}) and that ${\FRobI \not\subseteq \FReg}$ (Lemma~\ref{lemma:FRobInotsubsetFReg}). In Section~\ref{subsec:RobPreservationAfterVarInst} we will prove the upper bound $\FRob \not\subseteq \FRobI$ (Lemma~\ref{lemma:FRobnotsubsetFRobI}) and that $\FRobI \subseteq \FRob$ (Lemma~\ref{lemma:FRobIimpliesFRob}). Then, in Section \ref{subsec:InstVarVsAddEq}, we prove that $\FRobI \subseteq \FRobLEq$.
Finally, in Section~\ref{subsec:termination},  we prove Theorem \ref{thm:quasiquasidecidability}.

\subsection{Background on robustness and regularity}
\label{subsec:BackgroundRobustness}

In this section, we first give a formal definition of robustness, and then proceed to provide all the definitions needed to formally define our four classes of interest. 
We will also present some results regarding these definitions that will be used for proving the main theorem.

\

\emph{Notation}: Given a multivalued function $F:\Omega \subseteq \mathbb{R}^m\to \mathbb{R}^n$, we will denote with $F_1, \dots, F_n: \Omega \subseteq \mathbb{R}^m\to \mathbb{R}$ the univalued functions such that $F= (F_1, \dots, F_n)$. 
With a slight abuse of terminology, we will say that a function $F:\Omega \subseteq \mathbb{R}^m\to \mathbb{R}^n$ is \emph{satisfiable} 
if and only if it has a zero.


\subsubsection{Robustness.}
First, we provide a formal definition of the concept of robustness. Since our main focus are systems of equations, we will provide a definition of robustness only in terms of multivalued functions. This concept, however, can be generalized and formalized for general formulas. The definition that we give here is just a special case of the more general definition presented in  \cite{Franek:12}. In fact, a multivalued function $F$ is robustly satisfiable if and only if the logical formula representing the equation $F=0$ is. We will make use of the general definition of robustness only in the last section, when discussing Theorem \ref{thm:quasiquasidecidability}.

We first introduce the concept of distance between functions.

\begin{definition}[Distance between two functions]
	Let $F:\Omega_1 \subseteq \mathbb{R}^m\to \mathbb{R}^n$  and $F':\Omega_2 \subseteq \mathbb{R}^m\to \mathbb{R}^n$ be two multivalued continuous functions. 	Given $\Omega \subseteq \mathbb{R}^m$ such that $\Omega \subseteq \Omega_1$ and $\Omega \subseteq \Omega_2$,
	we define the distance between $F$ and $F'$ in $\Omega$ as
	$$\dist_{\Omega}(F,F') \defas \underset{k\in [1,n]}{\max}( \|F_k-F'_k\|_{\Omega} )$$	
	where $\|F_k-F'_k\|_{\Omega} \defas \sup \{|F_k({x})-F'_k({x})| : {x} \in {\Omega}\}$.
\end{definition}

When $\Omega$ is clear from the context, with an abuse of notation we will 
drop the subscript and just write $\dist(F,F')$. 
We say that $F'$ is an \emph{$\epsilon$-small perturbation} of $F$ if $\dist(F,F')< \epsilon$.

\begin{definition}[Robustness of a function]
	Given $\alpha\in \mathbb{R}_{> 0}$, we say that a continuous function $F:\Omega\subseteq \mathbb{R}^m\to \mathbb{R}^n$ is $\alpha$-robust iff for every continuous function $F'$ s.t. $\dist_{\Omega}(F,F') < \alpha$, either both $F$ and $F'$ have a zero in $\Omega$, or none of them has.
	A function $F$ is\textbf{ robust} iff there exists $\alpha\in \mathbb{R}_{> 0}$ s.t. $F$ is $\alpha$-robust in $\Omega'$.
	
\end{definition}

For $\Omega'\subseteq \Omega$, we say that $F$ is robust in $\Omega'$ iff $F_{|\Omega'}:\Omega'\subseteq \mathbb{R}^m\to \mathbb{R}^n$ is robust.

\begin{definition}[Robustly satisfiable function]
	\label{def:robustlysat}
	A function $F$ is \textbf{robustly satisfiable} iff
	it is robust and satisfiable.	
\end{definition}

\subsubsection{Robust solutions ($\FRob$).}
Robust satisfiability of a function---as defined by Definition~\ref{def:robustlysat}---does not depend on any specific solution. Indeed, it is the entirety of the solution set that accounts for the robust satisfiability of the function.  However, it will be useful to talk about the robustness of a function around a specific solution, which also formalizes the definition of the class $\FRob$.

\

	\begin{definition}[Robust solution]
		Given a function $F:\mathbb{R}^m\to \mathbb{R}^n$, we say that a point $p\in \mathbb{R}^m$ is a \textbf{robust solution} iff
		\begin{enumerate}
			\item $F(p)=0$, and
			\item for all $\epsilon>0$ there exists a $\delta > 0$ such that for all $F'$ with
			$\dist(F,F')<\delta$, there exists $p'$ such that 
			$F'(p')=0$ 
			and $\dist(p,p')<\epsilon$.
		\end{enumerate}
	\end{definition}
	
	It follows by the definition of robust solution that if $F$ has a robust solution, then $F$ is robustly sat. 
	
Note that the converse, however, is not true in general:

\begin{example}[Robustly sat formula with no robust solution]
	Let $F: \mathbb{R}\to \mathbb{R}$ defined by 
	$$F:x\mapsto
	\begin{cases}
		-x^2 & \text{ if } x<0 \\
		0        & \text{ if } 0 \leq x\leq 1 \\
		(x-1)^2        & \text{ if } 1<x\\
	\end{cases} $$
This function is robustly sat, since every $F'$ obtained by a small perturbation of $F$ has still a solution either near $0$ or near $1$. 
But no point in the solution set of $F$ (i.e. the points in $[0,1]$) is a robust solution.
Indeed, $0$ is not a robust solution, since, given $\epsilon = 0.1$, for every $\delta>0$, the function $F'_\delta$ defined by $F'_\delta(x)\mapsto F(x)-\delta$; 
for $0$ has no solution in the open ball $\openball{B}{0.1}(0)$. Symmetrically, for $1$ we can take $F''_\delta$ defined by $F''_\delta(x)\mapsto F(x)+\delta$. 
And for every point $p\in(0, 1)$, we can take $\epsilon=\min(\dist(p,0), \dist(p,1))$, and for every $\delta$ either $F'_\delta$ or $F''_\delta$.

\end{example}

\ 

A partial converse is the following result, which states that if a function is locally robustly satisfiable around a solution, the solution is robust.
\begin{proposition}
	\label{prop:locallyRobustlysat}
	If $p$ is a solution for $F$, and for every $\epsilon > 0$ there exists a neighborhood  $\Omega_\epsilon \subseteq \openball{B}{\epsilon}(p)$ of $p$ such that       
	$F$ is robustly sat in $\Omega_\epsilon$, then $p$ is a robust solution for $F$.
\end{proposition}

\begin{proof}
	For every $\epsilon>0$, $F$ is robustly sat in $\Omega_\epsilon$ if and only if  (by replacing the definition of robustly satisfiable function) $\forall \epsilon$, $\exists \delta$ s.t. $\forall F'$ with $\dist(F,F')<\delta$, $F'$ has a solution in $\Omega_\epsilon$.
	$F'$ has a solution in $\Omega_\epsilon$ if and only if there exists $p'$ s.t.  $F'(p')=0$ and $p'\in \Omega_\epsilon$ (i.e. $\dist(p, p')<\epsilon$).
	So we obtain: $\forall \epsilon$, $\exists \delta$ s.t. $\forall F'$ with $\dist(F,F')<\delta$, $\exists p'$ s.t. $p'$ is a solution for  $F'$ and $\dist(p, p')<\epsilon$ which is the definition of robust solution for $p$.
\end{proof}


\subsubsection{Robust under instantiation ($\FRobI$)}


In general, even if a solution is robust, after the instantiation of some variables, the projection of the solution may not be a robust solution for the function obtained after variable instantiation.

Now we provide some notation and some definitions regarding variable instantiation. We start by formalizing the fact that $k$ coordinates of a point have a finite representation in the form of a dyadic rational number.
\begin{definition}[$k$-finite point]
	A point ${p=(p_1,\cdots, p_m)\in \mathbb{R}^{m}}$ 
	is a \textbf{\kfinite{k}} 
	(with $0\leq k\leq m$)
	if for at least $k$ coordinates $i\in\{1,\dots,m\}$
 there exist integers $n_i$ and $r_i$ such that $p_i= n_i 2^{-{r_i}}$.
      \end{definition}
Here, we use base $2$ just for convenience. Any other base would work equally well for our purposes.    

	%

\begin{definition}[Robust instantiation of a point]
	Let ${F: B\subseteq \mathbb{R}^{n+k}\to \mathbb{R}^n}$ be a \Cone function  and $p=(p_1, \dots, p_{n+k})\in \mathbb{R}^{n+k}$ a \kfinite{k} (with $I$ denoting a set of  $k$ finitely representable indices).
	
	Given the partial assignment $\nu_I \defas \{x_i \mapsto p_i\}_{i \in I}$,
	we define the {\emph{instantiation of $F$ via $\nu_I$}} as the function
	$F_{|\nu_I}: B_{|\mathbb{R}^n} \subseteq \mathbb{R}^n\to \mathbb{R}^n$ 
        such that for every $(x_1,\dots,x_n)\in\mathbb{R}^n$,
        $F_{|\nu_I}(x_1,\dots,x_n)= F(y_1,\dots,y_{n+k})$, where 
                \[ \text{for } i\in \{1,\dots, n+k\},\:
          y_i=
          \left\{\begin{array}{l}
            p_i, \text{if } i\in I,\\
            x_i, \text{if } i\not\in I
          \end{array}\right.. \]

	We say that the partial assignment $\nu_I$ 
	is a \textbf{robust instantiation of $p$} if and only if the point $p_{|\nu_I} = (p_i)_{i\not\in I} \in \mathbb{R}^n$  
	 is a robust solution for $F_{|\nu_I}$.
      \end{definition}
      If an instantiation is not a robust instantiation, then we say it is a \emph{\mbox{non-robust} instantiation}. Note that assignments are defined as maps to the set of finitely representable values. So if $p$ is not a \kfinite{k}, then it admits no instantiations, and hence no robust instantiations.

For our method to succeed, we only need the existence of a single robust instantiation. 
So we are not interested in solutions that are robust under \emph{all} instantiations, but in solutions that are robust under \emph{at least one}  instantiation.
Finally, we have the following definition (from which the definition of $\FRobI$ follows):

\begin{definition}[Robust under instantiation] Given a continuous function ${F: B\subseteq \mathbb{R}^{n+k}\to \mathbb{R}^n}$ and a point $p$, we say that $p$ is \textbf{\mbox{robust under instantiation}}  if there exists at least one robust instantiation of $p$.
\end{definition}
If no robust instantiation of $p$ exists, then we say that $p$ is \emph{non-robust under instantiations}. If $p$ is not a \kfinite{k}, then, by definition, $p$ is non-robust under instantiations.

\

In the following, for ease of notation and without loss of generality, we will assume that ${I=\{n+1, \dots, n+k\}}$, unless otherwise specified.  
In this case, we will write 
$\nu$ instead of $\nu_I$, and we will denote
$\pn \defas (p_1 ,\dots, p_n)$ for the projection of $p$ to the first $n$ coordinates (i.e. the ones not instantiated by $\nu$), 
and $\pnk$ for the projection of $p$ to the last $k$ coordinates (i.e. the ones not instantiated by $\nu$).

\subsubsection{Robustness after equation adding ($\FRobLEq$).} 
We define $\FRobLEq$ as the set of \Cone functions $F: \mathbb{R}^{n+k}\to \mathbb{R}^{n}$ such that there exists a linear function $L: \mathbb{R}^{n+k}\to \mathbb{R}^k $ such that the function ${F_{leq}: \mathbb{R}^{n+k}\to \mathbb{R}^{n+k}}$, defined by $F_{leq}: x \mapsto (F(x), L(x))$,
has a robust solution.

\subsubsection{Regular solutions ($\FReg$).} We now define the last of our classes of interest. For doing so, we provide a brief background on the notion of regularity from the field of differential topology.

\begin{definition}[Regular point]
	Let $F: B\subseteq \mathbb{R}^m \to \mathbb{R}^n$ be a \Cone function, with $m-n = k \geq 0$. 
	We say that $p\in B$ is a \textbf{regular point} for $F$ 
	if and only if the Jacobian matrix of $F$ at $x$ has maximal rank.
\end{definition}
If $F(p)=0$ and $p$ is a regular point, we will say that $p$ is a \emph{regular solution} of $F$. 
If $p$ is not a regular point, we say that $p$ is a \emph{critical point}.  
We say that $q\in \mathbb{R}^n$ is a \emph{regular value} if and only if for every $p$ such that $F(p)=q$, $p$ is a regular point.
If $q$ is not a regular value, we say it is a \emph{critical value}.

Regularity is a very meaningful property, as it guarantees the "well-behavior" of the system. In particular, a regular point $p$ satisfies the hypothesis of the Implicit Function Theorem~\cite{Munkres1991AnalysisOM}, that we now recall and that we will use in Section \ref{subsec:GuaranteesRegularity}:

\emph{Implicit Function Theorem}: If  ${F(p_1 , \dots, p_n, p_{n+1}, \dots, p_{n+k}) = 0}$, and the Jacobian matrix of $F$ with respect to the first $n$ coordinates has non-zero determinant in $p$ (i.e. $\det(J_{F, x_{|n}}(p)) \neq 0$), 
then there exists a neighborhood $U\subseteq \mathbb{R}^k$ of $(p_{n+1}, \dots, p_{n+k})$ and a \Cone function $H: U \to \mathbb{R}^n$ such that $H(p_{n+1}, \dots, p_{n+k})=(p_1 , \dots, p_n)$, 
and such that, for all $q\in U$, ${F(H(q), q)=0}$. 


\subsection{Regularity and robustness under instantiation}
\label{subsec:GuaranteesRegularity}

In this section, we prove that the existence of a regular solution is a sufficient---but not necessary---criterion for the existence of a solution robust under instantiation.


  We prove that, if $F$ has at least one regular solution $q$, then there exists at least one regular solution $p$ such that at least $k$ coordinates of $p$ are finitely representable rational numbers, and such that the subsystem induced from the instantiation of the $k$ corresponding variables is robustly sat.

We first show that any regular solution is also a robust solution (Lemma \ref{lemma:regulImpliesRobust}), which will be useful to prove the main result of this section, Lemma \ref{thm:FRegsubsetFRob}, which implies  $\FReg \subseteq \FRobI$.  Then, we will show that this inclusion is strict by providing a counter-example in the form of a system of equations that has a solution robust under instantiation but no regular solutions (Lemma \ref{lemma:FRobInotsubsetFReg}).

\begin{lemma}
	\label{lemma:regulImpliesRobust}
	Given a \Cone function $F:B\subseteq\mathbb{R}^n\to \mathbb{R}^n$, if $p$ is a regular solution for $F$, then $p$ is a robust solution for $F$.
\end{lemma}

\begin{proof}
	Assume that $p$ is a regular solution of $F$. 
	Hence the Jacobian of $F$ at $p$ has maximal rank. We prove that $p$ is a robust solution of $F$. 
	So let $\varepsilon>0$ be arbitrary, but fixed
	and such that $p$ is the unique solution of $F$ in $\openball{B}{\epsilon}(p)$.
	Such $\varepsilon$ always exists, 
        since by the inverse function theorem $F$ maps a neighborhood of $p$ diffeomorphically onto an open set of $\mathbb{R}^n$~\cite[Chapter 1.2]{Milnor1965TopologyFT}.
	Hence $0\not\in \partial \openball{B}{\epsilon}(p)$, and, since $p$ is the only solution of $F$ in   $\openball{B}{\epsilon}(p)$, and it is regular, then, by definition,
	$\deg(F, \openball{B}{\epsilon}(p), 0) = |\det(J_F(p))| \neq 0$.	
	Let $\delta<\min_{x\in\partial \openball{B}{\epsilon}(p)} |F(x)|$. 
	 By Lemma~1~\cite{Franek:12}, every $F'$ with $\dist(F, F')<\delta$ has a zero in $\openball{B}{\epsilon}(p)$. 
	This proves that $p$ is a robust solution for $F$.
\end{proof}

\begin{lemma}
	\label{thm:FRegsubsetFRob}
	Let $F:B\subseteq\mathbb{R}^{m}\to \mathbb{R}^n$ (with $m=n+k$) be a \Cone function. If there exists a regular solution $q$ of $F$, then there exists a regular solution $p$ of $F$  in a neighborhood of $q$ such that $p$ is robust under instantiation.
	
\end{lemma} 

\begin{proof}
	
	If $q$ is a regular solution for $F$, then $J_F(q)$ has maximum rank. 
	Since a rectangular matrix has maximum rank if and only if one of its maximal square sub-matrix has maximum rank,
	 then, without loss of generality, we can reorder the variables so that the square sub-matrix given by the first $n$ columns has maximum rank, 
	 i.e. $\det(J_{F, \projn{x}}(q)) \neq 0$. 
	By the Implicit Function Theorem, there exists a neighborhood $U\subseteq \mathbb{R}^k$ of $(q_{n+1}, \dots, q_{n+k})$ and a \Cone function ${H: U \to \mathbb{R}^n}$ such that $H(q_{n+1}, \dots, q_{n+k})=(q_1 , \dots, q_n)$, and such that, for all $q'\in U$, ${F(H(q'), q')=0}$.
	
	In general, it is not guaranteed that every $(H(q'), q')$ will be a regular point for $F$. 
	However, since the Jacobian $J_F: \mathbb{R}^m \to \mathbb{R}^{m \times n}$%
	, the projection ${\pi_{n \times n}: \mathbb{R}^{m \times n} \to \mathbb{R}^{n \times n}}$
	(that projects a $m\times n$ matrix onto the $n\times n$ sub-matrix of its first $n$ columns)
	and the determinant $\det: \mathbb{R}^{n \times n} \to \mathbb{R}$ are all continuous functions, then the set $U_r \subseteq \mathbb{R}^m$, 
	consisting of all the  points $q'$ for which $\det(J_{F,\projn{x}}(q'))\neq 0$, is open, since  $\mathbb{R} \setminus \{0\}$ is open and, by definition, $U_r = (det \circ \pi_{n \times n} \circ J_F)^{-1}(\mathbb{R} \setminus \{0\})$.
	
	Let us consider the projection of $U_r$ over $\mathbb{R}^k$, i.e. $U_{r_{|k}} = \pi_k(U_r) = \{(q'_{n+1}, \dots, q'_{n+k}) \in \mathbb{R}^k | (q'_1 , \dots, q'_n, q'_{n+1}, \dots, q'_{n+k}) \in U_r  \}$. Since projections are open maps, then $U_{r_{|k}}$ is open.  Since both $U$ and $U_{r_{|k}}$ are neighborhoods of $(q_{n+1}, \dots, q_{n+k})$, then their intersection $U'\defas U \cap U_{r_{|k}}$ is again a neighborhood of $(q_{n+1}, \dots, q_{n+k})$.

	Now we prove that $U'$ contains at least one  \kfinite{k} $p'$. 
	The set of \kfinites{k} in $\mathbb{R}^k$ is exactly the set of points having coordinates with finite representation.
	Let us call this set $A$.  
	We have that $A$ is dense in $\mathbb{R}^k$, as, for every point ${z=(z_1,\dots , z_k) \in  \mathbb{R}^k}$, $z$ is the limit of the sequence $\{([z_1]_i, \dots , [z_k]_i) \}_{i\in \mathbb{N}} \subseteq A$, where $[z_j]_i$ is the truncation of $z_j$ to the $i$-th digit after the zero. 
	Since $A$ is dense in $\mathbb{R}^k$, then $A$ intersects every non-empty open of $\mathbb{R}^k$. 
	In particular $A \cap U' \neq \emptyset$, hence there exists $p'\in A \cap U'$. 
	
	Let $p\defas(H(p'), p')\in \mathbb{R}^m$. 
	We have that $p$ is a solution for $F$ (since $F(p) = F(H(p'), p')=0$). 
	Moreover, $p$ is also regular, since $p'\in U' \subseteq U_{r_{|k}}$ and hence $(H(p'), p') \in U_r$.
	
	Let $\nu \defas \{x_i \mapsto p_i\}_{i \in [n+1,  n+k]}$. 
	We have that the point $\pn = H(p') \in \mathbb{R}^n$ is a regular point for $\Fv$. Indeed, since $p$ is a regular point, and $J_{F, x_{|n}}(p)$ depends only on the first $n$ coordinates, then $\det(J_{\Fv}(\pn)) \neq 0$.
	
	Since $\pn$ is a regular solution for $\Fv$,  by  Lemma~\ref{lemma:regulImpliesRobust}, $\pn$ is also a robust solution for $\Fv$. Hence $p$ is robust under instantiation.
\end{proof}

Modifying the proof by choosing directly  $p$ as $q$ we get:
\begin{corollary}
	\label{corollary:kfiniteAndRegImpliesRobI}	
	Let $F:B\subseteq\mathbb{R}^{m}\to \mathbb{R}^n$ (with $m=n+k$) be a \Cone function
	  If $q$ is  both a regular solution and a \kfinite{k}, then $q$ is robust under instantiation.
\end{corollary}

Now we show that the converse of Lemma \ref{thm:FRegsubsetFRob} does not hold, i.e. that the existence of a solution robust under instantiation does not imply the existence of a regular solution. Consider the following example:
\begin{example}[Critical solution, but robust under instantiation]
	Let $F: [0, 1]^2 \subseteq \mathbb{R}^2\to \mathbb{R}$ defined by $F(x,y) = (x^2-y^3)$,
	and let $p = (0, 0)$. $J_F(p)=(0,0)$ has non-maximum rank (hence $p$ is a critical solution), but the instantiation $\{x\mapsto 0 \}$ leads to the subsystem $-y^3 = 0$, which is robust.
\end{example}

In this example, we could have chosen a different point, say $p'=(1,1)$, which is both robust under instantiation and regular. 

However, this is not always possible. A system can have a solution robust under instantiation, but no regular solutions. Indeed:

\begin{lemma}
	\label{lemma:FRobInotsubsetFReg}
	$\FRobI \not\subseteq \FReg$
\end{lemma}
\begin{proof}
	Let ${F:[-1,1]^3\subseteq \mathbb{R}^3\to \mathbb{R}^2}$ defined by 
	\begin{equation*}
		F(x_1, x_2, x_3)=
		\begin{cases}
			x_1^3 & (F_1)\\
			x_2+x_3 & (F_2)
		\end{cases}
	\end{equation*}
	
	\ \\
	The point $(0,0,0)$ is robust under instantiation. Indeed, the instantiation ${\{x_3 \mapsto 0 \}}$ leads to the following system of equations
	\begin{equation*}
		F'(x_1, x_2)=
		\begin{cases}
			x_1^3 & (F'_1)\\
			x_2 & (F'_2)
		\end{cases}
	\end{equation*}
	which has non-zero degree, hence it is robustly sat. 
	It is easy to show that the degree of $F'$ in $[-1,1]^2$ is non-zero. 
	In fact, $F'_1$ depends only on $x_1$ and $F'_2$ only on $x_2$,
	and for both $F'_1$ and $F'_2$ it suffices to apply the Intermediate Value Theorem 
	to prove that $\deg(F'_i, [-1,1], 0) \neq 0$ (for $i=1,2$).
	Since the degree of the Cartesian product is the product of the degrees~\cite[Theorem 7.1.1]{BrouwerDegreeDincaMawhin},
	$\deg(F', [-1,1]^2, 0) = \deg(F'_1, [-1,1], 0) * \deg(F'_2, [-1,1], 0)  \neq 0 $.


	So $F\in \FRobI$.	
	However, $F$ does not have any regular solution. 
	Indeed, the first equation implies that for every every solution  $p$ its first coordinate has to be $p_1 = 0$.
	Since, for such a solution $p$, the first row of $J_F(p)$ is everywhere $0$, then the Jacobian cannot have maximum rank. 
	Hence every solution $p$ is not regular, i.e. $F\not\in \FReg$.
\end{proof}

\subsection{Robustness preservation after variable instantiation}
\label{subsec:RobPreservationAfterVarInst}

In the previous section, we have proven that, under the  assumption of the existence of a regular solution, there exists a solution that is robust under instantiation.

But what happens if we drop the assumption of regularity? In general, if we don't put any restriction on the functions we are considering, the solution space can be arbitrarily complicated. 
Indeed, for every closed subset $K \subseteq \mathbb{R}^m$, there exists a smooth function $F$ such that $F^{-1}(0)=K$ (\cite[Theorem 2.29]{LeeIntroSmooth}).

One may hope that, by restricting to functions that have a robust solution, we can always find a solution robust under instantiation. In this section, we show that this, unfortunately, does not hold. Consider the following example.
\begin{example}[Robust solution, but non-robust under instantiations]
	\label{ex:robButNonrobUnderInst}
	Let ${F: [-1,1]^2\subseteq \mathbb{R}^2\to \mathbb{R}}$ defined by $F(x,y) = (x^2 - y^2)$,
	and let $p = (0, 0)$. It is easy to show that $p$ is a robust solution. However, whether we instantiate $\{x\mapsto0\}$ or $\{y\mapsto 0\}$, 
	the resulting subfunctions (resp. $F_{|\{x\mapsto0\}}(y)=y^2$ and $F_{|\{y\mapsto0\}}(x)=x^2$)  are not robust.
\end{example}
In this example, we could have chosen another point, for example ${p'=(1,1)}$, which is regular (since $J_F(p')=(2,-2)$), and hence, by Corollary \ref{corollary:kfiniteAndRegImpliesRobI}, robust under instantiation. But in general, this is not always possible. Indeed, we have the following result:

\begin{lemma}
	\label{lemma:FRobnotsubsetFRobI}
	$\FRob \not\subseteq \FRobI$
\end{lemma}
\begin{proof}
	\label{ex:robNotRobI}
	Let $F: [-1, 1]^4\subseteq \mathbb{R}^4 \to \mathbb{R}^3$ defined by 
	
	\begin{equation*}
		F(x_1, x_2, x_3, x_4)=
		\begin{cases}
			x_1^2+x_2^2-x_3^2-x_4^2 \\
			
			2(x_1 x_4+x_2 x_3) \\
			
			2(x_2 x_4-x_1 x_3)  \\
		\end{cases}
	\end{equation*}
	
	It is easy to show that $p=(0,0,0,0)$ is the only solution of $F$. Moreover, $p$ is robust (see the discussion about Hopf maps in 
	\cite{FranekHopf})
	, hence $F\in \FRob$. However, no instantiation $\nu_i \defas \{x_i \mapsto 0\}$ is robust. Indeed,  $(0, 0, 0)$ is the only solution of $F_{|\nu_i}$ in $[-1,1]^3$, but $\deg(F_{|\nu_i}, [-1,1]^3, 0) = 0$ (remember that, if a system of equations $F$ has an isolated robust solution in $B$, then $\deg(F, B, 0)\neq 0$). So $F\not\in \FRobI$.
\end{proof}


Now we show that the converse holds, i.e. that  every solution robust under instantiation is robust:

\begin{lemma}
	\label{lemma:FRobIimpliesFRob}
	Let $F:B\subseteq \mathbb{R}^{n+k}\to \mathbb{R}^n$. If $p$ is a solution robust under instantiation, then $p$ is a robust solution. 
\end{lemma}

\begin{proof}
	If $p$ is a solution robust under instantiation, then there exists a set of indices $I$ (w.l.o.g. say $I=\{n+1, \dots, n+k\}$) 
	and a corresponding instantiation $\nu$ such that $\pn$ is a robust solution for $\Fv : B_{|\mathbb{R}^n} \subseteq \mathbb{R}^n\to \mathbb{R}^n$, that is,
	 for every $\epsilon>0$ there is a $\delta$ such that for all $\Fv'$ with $\dist(\Fv,\Fv')<\delta$, 
	there exists a solution $\pn'$ of $\Fv'$ with $\dist(\pn, \pn')<\epsilon$.

	To prove that $p$ is a robust solution for $F$ let $\epsilon>0$ be arbitrary, but fixed, and take the corresponding $\delta$ as ensured by robustness under instantiation.
	For every $F'$ with $\dist(F,F')<\delta$,
	we have that also $\dist(\Fv,
	\Fv')<\delta$, hence there exists $\pn'$ that satisfies $\Fv'$ with $\dist(\pn, \pn')<\epsilon$. If $\pn'$ satisfies $\Fv'$, then $p'\defas (\pn', \pnk)$ satisfies $F'$. Since $\dist(p, p') = \dist(\pn, \pn')< \epsilon$,  $p$ satisfies the definition of robust solution for~$F$.
\end{proof}

A straightforward corollary of Lemma \ref{lemma:FRobIimpliesFRob}, is that  $\FRobI \subseteq \FRob$
which
concludes the proof that $\FRobI \mysubsetneq \FRob$.  Together with Lemma~\ref{lemma:FRobInotsubsetFReg} and Lemma~\ref{thm:FRegsubsetFRob}, this implies Theorem~\ref{thm:FRobIbounds}.

\subsection{Variable instantiation vs. equation adding}
\label{subsec:InstVarVsAddEq}

Given an under-constrained system of equations, we showed that there are two different ways to reduce to a well-constrained system of equations: decreasing the number of variables (i.e. instantiations) or increasing the number equations. In this section we will show that the class of systems that can be solved via variable instantiation is a subset of the class of systems that can be solved via adding equations.

Recall our definition of $\FRobLEq$ as the set of functions $F: B \subseteq \mathbb{R}^{n+k}\to \mathbb{R}^{n}$ such that there exists a linear function $L: \mathbb{R}^{n+k}\to \mathbb{R}^k $ such that the function ${F_{leq}: B \subseteq \mathbb{R}^{n+k}\to \mathbb{R}^{n+k}}$, defined by $F_{leq}: x \mapsto (F(x), L(x))$,
 has a robust solution.


Given any partial assignment $\nu \defas \{x_i \mapsto p_i\}_{i \in [n+1, n+k]}$, 
we can consider the function $F_{leq} = (F, L)$, given by $F$ and by the linear function ${L: B\subseteq \mathbb{R}^{n+k}\to \mathbb{R}^k}$
defined by ${L:(x_1,\dots, x_{n+k})\mapsto (x_{n+1} - p_{n+1}, \dots, x_{n+k}-p_{n+k})}$. Equivalently, since $L$ only depends on the last $k$ variables, we can consider it as a function ${L: B_{|\mathbb{R}^k}\subseteq \mathbb{R}^k\to \mathbb{R}^k}$. 

 
 While it is trivial to show that every solution of $F_{|\nu}$  is also a solution of $F_{leq}$, we need to prove that also the robustness of a solution is preserved. 

\begin{lemma}
	\label{lemma:instVarVsAddEq}
	Given a system of equations $F = 0$ (with $F: B \subseteq \mathbb{R}^{n+k}\to \mathbb{R}^n$),
	a point $p=(p_1, \dots, p_{n+k}) \in B$ and subset of indexes $I:=\{n+1 , n+2, \dots, n+k\}$, let us consider the following statements:
	
	\begin{enumerate}
		\item For the function $\Fv:  B_{|\mathbb{R}^n} \subseteq \mathbb{R}^n\to \mathbb{R}^n$, obtained by instantiating variables via $\nu = \{x_i \mapsto p_i\}_{i\in I}$, the point $\pn \defas (p_1,\dots, p_n)$ is a robust solution 
		\item For the function $F_{leq}:  B \subseteq \mathbb{R}^{n+k}\to \mathbb{R}^{n+k}$, defined by
		$$F_{leq}(x_1, \dots, x_{n+k}) \mapsto (F(x_1, \dots, x_{n+k}), L(x_{n+1}, \dots, x_{n+k}))$$ where
		$L: B_{|\mathbb{R}^k} \subseteq  \mathbb{R}^k\to \mathbb{R}^k$ is defined by \[{L:(x_{n+1},\dots, x_{n+k})\mapsto (x_{n+1} - p_{n+1}, \dots, x_{n+k}-p_{n+k})},\] the point $p$ is a robust solution.
	\end{enumerate}
	\ 
Then, it holds that $1.$ implies $2.$.
\end{lemma}

\begin{proof}

	To prove that
	1. $\Rightarrow$ 2., it will be useful to consider a third auxiliary condition:
	\begin{enumerate}
		\item[\emph{3.}] 
		\emph{For the function $F_{{|\nu}_{leq}} \defas F_{|\nu} \times L : B \subseteq \mathbb{R}^{n+k}\to \mathbb{R}^{n+k}$ obtained by the Cartesian product between $F_{|\nu}$ and the linear function $L$, the point $p$ is a robust solution
	}
	\end{enumerate}
	and prove first that 
	$1. \Rightarrow 3.$ and then that  
	$3. \Rightarrow 2.$
	 
	 ($1. \Rightarrow 3.$ )
	By Theorem 7.1.1 \cite{BrouwerDegreeDincaMawhin}, the degree of the Cartesian product is the product of the degrees,
	i.e., for every $\Omega = \Omega_1 \times \Omega_2 \subseteq \mathbb{R}^{n}\times \mathbb{R}^k$,
	\begin{equation}
		\deg(F_{{|\nu}_{leq}}, \Omega, 0) = \deg(F_{|\nu}, \Omega_1, 0) * \deg(L, \Omega_2, 0 ) 
	\end{equation}
	Since $J_L(\pnk)$ is the identity matrix, the point $\pnk$ is a regular solution for $L$. Since $\pnk$ is the only solution of $L$, then, for every $\Omega_2 \subseteq \mathbb{R}^{k}$ such that $\pnk \in \Omega_2$, $\deg(L, \Omega_2, 0 ) = \det(J_L(\pnk)) = 1$.
	  
	Thus, for every $\Omega = \Omega_1 \times \Omega_2 \subseteq \mathbb{R}^{n}\times \mathbb{R}^k$ such that $p\in \Omega$, we have that
	\begin{equation}
		\label{eq:Thm711}
		\deg(F_{{|\nu}_{leq}}, \Omega, 0)  = \deg(F_{|\nu}, \Omega_1, 0)
	\end{equation}

	 If 1. holds, then,  by Thm. 6 \cite{Franek:12}, for every $\epsilon>0$ there exists an open ${\Omega_{1,\epsilon} \subseteq \openball{B}{\epsilon}(\pn)}$ such that $\deg(F_{|\nu}, \Omega_{1,\epsilon}, 0) \neq 0$. 
	 Let 
	 $\Omega_2$ be any neighborhood of $\pnk$ 
	 s.t. $\Omega_2 \subseteq \openball{B}{\epsilon}(\pnk)$,
	 and let $\Omega_\epsilon \defas \Omega_{1,\epsilon} \times \Omega_2$. 	 
	 By Equation \ref{eq:Thm711}, ${\deg(F_{{|\nu}_{leq}}, \Omega_\epsilon, 0) \neq 0}$. 
	 Hence,
	 for every $\epsilon>0$, $F_{{|\nu}_{leq}}$ is robustly sat in $ \Omega_{1,\epsilon}$. 
	 So we have constructed, for every $\epsilon>0$, a neighborhood $\Omega_{1,\epsilon} \subseteq \openball{B}{\epsilon}(p)_{\epsilon}$ of $p$ in which $F_{{|\nu}_{leq}}$ is robustly sat.  
	 By Proposition \ref{prop:locallyRobustlysat}, this implies that $p$ is a robust solution for $F_{{|\nu}_{leq}}$.

	 \ 
	 
	($3. \Rightarrow 2.$ )
	Now we show that if $p$  is a robust solution for $F_{{|\nu}_{leq}}$ then it is a robust solution for    $F_{leq}$.	We will construct a homotopy between $F_{{|\nu}_{leq}}$  and    $F_{leq}$, and then rely on the Homotopy Invariance Property of the topological degree to prove the claim.
	
	Let $S:  \mathbb{R}^{n+k }\times [0,1] \to \mathbb{R}^{n+k }$ be defined by
        \vspace*{-0.32cm}\begin{multline*}
          S:((x_1, \dots, x_n, x_{n+1},\dots x_{n+k}), t) \mapsto\\
              (x_1, \dots, x_n, \ tx_{n+1} + (1-t)p_{n+1},\ \dots \ ,\ tx_{n+k} + (1-t)p_{n+k})
        \end{multline*}          
	
	The map $H: B \times [0,1]  \subseteq   \mathbb{R}^{n+k} \times [0,1]  \to \mathbb{R}^{n+k}$ 
	defined by
	$$H:(x_1, \dots, x_{n+k}, t)\mapsto (( F\circ S)(x_1, \dots, x_{n+k}, t) \ , \ L((x_{n+1}, \dots, x_{n+k}))) $$
	 is a homotopy between $F_{{|\nu}_{leq}}$ and $F_{leq}$ since
	 $H(\cdot, 0) \equiv F_{{|\nu}_{leq}}$, $H(\cdot, 1) \equiv F_{leq}$, and $H$ is continuous, being the composition of continuous functions. 
	
	It is easy to see that the functions $F_{leq}$ and $F_{{|\nu}_{leq}}$ have exactly the same solution space. Indeed, every solution of $F_{leq}$ has to satisfy the equations given by $L$, i.e. ${x_{n+1}=p_{n+1}, \dots, x_{n+k}=p_{n+k}}$. 
	By replacing in $F_{leq}$  every $x_i$ with $p_i$, for ${i\in [n+1, n+k]}$, we obtain exactly $F_{{|\nu}_{leq}}$. 
	
	Furthermore, for every $t\in [0,1]$ the solution space of $H(\cdot , t)$ is the same as $H(\cdot , 0) \equiv F_{{|\nu}_{leq}}$. 
	Indeed, for every $t$, a solution of $H(\cdot , t)$ has to satisfy the equations given by $L.$ 
	Then, by replacing  every $x_i$ with $p_i$ for $i\in[n+1, n+k]$, we have that every $tx_i + (1-t)p_i $ is replaced by $tp_i + (1-t)p_i$, which is equal to $p_i$. Thus we reduced  again to  $F_{{|\nu}_{leq}}$.
	
	
	Now, suppose that 3. holds. Then, for every $\epsilon>0$, there exists $\Omega_\epsilon \subseteq \openball{B}{\epsilon}(p) $ such that $\deg(F_{{|\nu}_{leq}}, \Omega_\epsilon, 0)\neq 0$. 
	This implies that  ${0\not \in F_{{|\nu}_{leq}}(\partial  \Omega_\epsilon)}$, hence  ${0\not \in H(\partial  \Omega_\epsilon, 1)}$.
	 So, by the previous observation, $0\not \in H(\partial  \Omega_\epsilon, t)$ for every $t\in [0,1]$. Hence $0\not \in H(\partial  \Omega_\epsilon, [0,1])$, and we can apply the Homotopy Invariance Property.
	
	The Homotopy Invariance Property of the topological degree states that, if $0\not\in H(\partial  \Omega_\epsilon \times [0,1])$, then $\deg(H(\cdot, 0),  \Omega_\epsilon, 0) = \deg(H(\cdot, 1),  \Omega_\epsilon, 0)$, i.e. ${\deg(F_{{|\nu}_{leq}},  \Omega_\epsilon, 0) = \deg(F_{leq},  \Omega_\epsilon, 0)}$. 
	So we have constructed, for every $\epsilon >0$, a neighborhood $\Omega_\epsilon \subseteq \openball{B}{\epsilon}(p)$ of $p$ in which $F_{leq}$ is robustly sat. By Proposition~\ref{prop:locallyRobustlysat}, this implies that $p$ is a robust solution for  $F_{leq}$.
	
\end{proof}

So robustness of the system obtained by variable instantiation implies robustness of  the system obtained by adding the equalities corresponding to this variable instantiation.
Theorem~\ref{thm:instVersusEq} is a straight-forward consequence.

\subsection{Termination}
\label{subsec:termination}
The method discussed in Section \ref{sec:certificate-search} made use of numerical optimization to enumerate the points over which the variable instantiation method is applied. This technique, while practically very efficient---as shown by our experiments---is not guaranteed to terminate, in general. Indeed, even in the bounded case, numerical optimization does not guarantee full coverage of the space.

In this section, we present a variation of our method that uses a different technique for enumerating points, and that is guaranteed to terminate on problems in $\FRobI$. While this variation is not intended to be of practical use, it will serve the purpose of proving Theorem \ref{thm:quasiquasidecidability}.

\ 

Given $F: B \subseteq \mathbb{R}^{n+k} \to \mathbb{R}^n$, if $F \in \FRobI$, by definition we know that there exists a \kfinite{k} $p$ that is robust under instantiation. 
Such a point~$p$ is, in general, not a \kfinite{(n+k)} (indeed,  there are problems for which no solution is a \kfinite{(n+k)}).
Hence no point enumeration technique is guaranteed to find precisely $p$, 
since only \kfinites{(n+k)}  can be expressed explicitly. 
However, this is not an actual limitation. 
Indeed, for our method to succeed, we don't necessarily need to explicitly produce a solution. We just need to find a point sufficiently close to an actual solution, and that shares with the solution $k$ indices, so that, after the instantiation of the corresponding $k$ variables, we end up with a subproblem that is robustly satisfiable. This suffices to produce a certificate.

The following lemma shows that, for every problem in $\FRobI$, it is always possible to find a \kfinite{(n+k)} and a partial assignment such that the resulting subproblem is robustly satisfiable.

\begin{lemma}
	\label{lemma:nkfinite}
	 Let $F : B \subseteq \mathbb{R}^{n+k} \to \mathbb{R}^n$. If $F \in \FRobI$, then there exist a {\kfinite{(n+k)}}  $p'\in B$ and a partial assignment $\nu' \defas \{x_i \to p'_i\}_{i\in I}$ (with $I$ being a set of $k$ indices), such that the function ${F_{|\nu'}: B_{|\mathbb{R}^n} \subseteq \mathbb{R}^n \to \mathbb{R}^n}$ is robustly satisfiable.
 \end{lemma}
\begin{proof}
	$F \in \FRobI$ means that there exists a \kfinite{k} $p$ 
	(w.l.o.g. say the $k$ finitely representable indices are $[n+1, n+k]$) 
	that is robust under instantiation,
	i.e. there exists a partial assignment $\nu\defas \{x_i \mapsto p_i\}_{i\in [n+1, n+k]}$ 
	such that $p_{|\nu}$ is a robust solution for $\Fv$. 
	This implies that $\Fv$ is robustly satisfiable in  $B_{|\mathbb{R}^n}$.
	
	Now, given any $p' \in B$ such that $p'_{n+1} = p_{n+1}, \dots, p'_{n+k} = p_{n+k}$, 
	and, given $\nu'\defas \{x_i \mapsto p'_i\}_{i\in [n+1, n+k]}$, 
	we have that $\nu' \equiv \nu$, hence $\Fv' \equiv F_{|\nu}$, 
	which implies that $F_{|\nu'}$ is robustly satisfiable in  $B_{|\mathbb{R}^n}$. 
	In order to find a $p'$ that respects the statement conditions, first we fix the last $k$ coordinates to be equal to $(p_{n+1},\dots,p_{n+k})$. 
	Then, since the set of \kfinites{n} is dense in $\mathbb{R}^{n}$, and hence intersects $\interior{B_{|\mathbb{R}^n}}$ (being an open), 
	there exists a \kfinite{n} $(p'_1, \dots, p'_n) \in \interior{B_{|\mathbb{R}^n}}$.
	If we take such point, and append the last $k$ coordinates previously fixed, 
	we obtain a point $p'\defas (p'_1, \dots, p'_n, p_{n+1}, \dots, p_{n+k})$. Since the last $k$ coordinates of $p'$ coincides with the last $k$ coordinates of $p$, we have that $F_{|\nu'}$ is robustly satisfiable in  $B_{|\mathbb{R}^n}$. 
	Moreover, such $p'$ is a \kfinite{(n+k)}. 
	Indeed, the part consisting of the first $n$ coordinates is $n$-finite by construction, 
	while the second part consisting of the last $k$ coordinates is $k$-finite because $p_{n+1},\dots,p_{n+k}$ is.
\end{proof}

Considering a bounded system of equations and inequalities satisfiable iff it is satisfiable by
a variable assignment the assigns values within the corresponding interval to all variables,
it is straightforward to extend the definitions from Section~\ref{subsec:BackgroundRobustness} analogically from systems of equations to bounded systems of equations and inequalities. Based on this, we can prove the following theorem. 
\begin{theorem}
	\label{thm:quasiquasidecidability}
	There exists a procedure that, given a bounded system of equations and inequalities $F=0 \wedge G \leq 0$,
	\begin{itemize}
        \item always returns the correct answer ``satisfiable'' or ``unsatisfiable'', if it terminates,
        \item always terminates successfully when $F=0 \wedge G \leq 0$ is robustly satisfiable  and  $F\in \FRobI$,
        \item always terminates successfully when $F=0 \wedge G \leq 0$ is robustly unsatisfiable.
	\end{itemize}
\end{theorem}

	\begin{proof}
	 
	We first concentrate on the second point, by showing a procedure that always correctly terminates on problems in $\FRobI$.

By Lemma \ref{lemma:nkfinite}, we have that,
for every $F: B \subseteq \mathbb{R}^{n+k}\to \mathbb{R}^n $ such that $F \in \FRobI$, there exists a \kfinite{(n+k)}~$p'$ and a partial assignment $\nu'$ such that $F_{|\nu'}$ is robustly satisfiable.
We can always find such $p'$ and $\nu'$. 
Indeed,
since the set of \kfinites{(n+k)} is countable, we can construct a well-order: 
say, for example, we first take the finite set of points whose  coordinates are represented by at most $1$ digit 
(and sort it by lexicographic order),
then the finite set of points  whose coordinates are represented by at most $2$ digits, and so on
\footnote{Note that this is independent by the Axiom of Choice, which is needed only in the case of uncountable sets.}.
For each such point $p$, and for each instantiation $\nu$ (note that the set of possible instantiations is finite), we can consider the   
resulting subsystem $F_{|\nu'} : B_{|\mathbb{R}^n}\subseteq \mathbb{R}^n \to \mathbb{R}^n$, 
and then apply the box-gridding procedure
discussed in Section \ref{subsec:box}, which---without the stopping criterion---is guaranteed to terminate on robust instances~\cite{Franek:12}. Note that box-gridding method also handles inequalities.

The procedure described so far is not yet guaranteed to converge. 
Indeed, while box-gridding is guaranteed to terminate on robust instances, it could diverge on non-robust instances, thus preventing the general procedure to terminate (either because one point yielded
before $p'$ was non-robust, or because one instantiation tried before $\nu'$ was a non-robust instantiation).

We can overcome this problem by using, instead of depth-first search, a technique called dove-tailing. 
Indeed, we have an infinite sequence of problems (given by the combination of points and instantiations), and, for each, a (possibly infinite) sequence of box-gridding iterations. 
We outline the following iterative procedure:
\begin{itemize}
	\item For $i=1$, we perform the $1$-st box-gridding iteration on the $1$-st problem.
	\item For $i=2$, we perform the $2$-nd box-gridding iteration on the first problem, and then the $1$-st box-gridding iteration on the second problem.
	\item \dots
	\item For $i=N$, we perform the $N$-th box-gridding iteration on the first problem, then the $(N-1)$-th box-gridding iteration on the second problem, \dots, and then the $1$-st box-gridding iteration on the $N$-th problem.	
\end{itemize}
First, we are guaranteed to find the problem given by the point $p'$ and the variable instantiation $\nu'$  after a finite amount of steps. Given such problem, we are guaranteed that also the box-gridding procedure will terminate after finitely many iterations. Hence, also our general procedure is guaranteed to terminate. 
\ 

The third point regarding robustly unsatisfiable problems simply follows by the use of box-gridding on the whole system using the bounds of the given system of equations and inequations as its starting box. Indeed, if the system is robustly unsatisfiable, then this  is guaranteed to terminate with a correct result.

To finalize the proof of the theorem, it suffices to consider the procedure that runs  the two previous procedures in parallel.

\end{proof}

Note that for formulas of the given form (bounded system of equations and inequalities), the procedure described in the proof of Theorem~\ref{thm:quasiquasidecidability} can be seen as an instantiation of the certificate search method presented in Section~\ref{sec:method} that uses exhaustive enumeration on the level of points and instantiations, and directly uses the given bounds as the starting box for box gridding. The theorem shows that 
under robustness assumptions, 
such an instantiation will always terminate successfully for bounded system of equations and inequalities. However, complete enumeration makes this instantiation hopelessly inefficient in practice, and goal oriented methods, as discussed in the first part of the paper, are necessary for practical efficiency. Also, there is no known way of algorithmically deciding whether a given formula satisfies the robustness precondition that ensures termination of the procedure, and hence Theorem~\ref{thm:quasiquasidecidability} is \emph{not} a decidability result.


 \section{Related Work}
\label{sec:relatedwork}

The computation of certificates for formulas \emph{not} being satisfiable in various first-order theories has been an important research topic of the SAT modulo theory community~\cite{Barbosa:23} over recent years. In the case of satisfiable formulas, this topic has---to our knowledge---been restricted to the \smtnta, since for most other theories used in an SMT context, satisfying assignments have a straight-forward representation.

One strategy for proving satisfiability in \smtnta is to prove a stricter requirement that implies satisfiability, but is easier to check. For example, one can prove that \emph{all} elements of a set of variable assignments satisfy the given formula~\cite{iSAT3}, or that a given variable assignment satisfies the formula for \emph{all possible interpretations} of the involved transcendental functions within some bounds~\cite{incrlin}. Such methods may be quite efficient in proving satisfiability of formulas with inequalities only, since those often have full-dimensional solution sets. However, such methods usually fail to prove satisfiability of equalities, except for special cases with straightforward rational solutions.

Computation of formally verified solutions of square systems of equations is a classical topic in the area of interval analysis~\cite{Rump:10,Neumaier:90,Moore:09}. Such methods usually reduce the problem either to fixpoint theorems such as Brouwer's fixpoint theorem or special cases of the topological degree, for example, Miranda's theorem. Such tests are easier to implement, but less powerful than the topological degree (the former fails to verify equalities with double roots, such as $x^3=0$, and the latter requires the solution sets of the individual equalities to roughly lie normal to the axes of the coordinate system).

In the area of rigorous global optimization, such techniques are applied~\cite{Hansen:92,Kearfott:98} to conjunctions of equalities and inequalities in a similar way as in this paper, but with a slightly different goal: to compute rigorous upper bounds on the global minimum of an optimization problem. This minimum is often attained at the boundary of the solution set of the given inequalities, whereas satisfiability is typically easier to prove far away from this boundary.

There are several fragments of \nta for which real root isolation methods have appeared~\cite{StrzeboskiExpLogArctan, MCCALLUM201216,StrzebonskiTame,RealRootIsolPolyPower,RootIsolMixedTrigPol, ReductionTranscDecProb}.
However, those fragments only allow univariate functions, they are restricted to certain transcendental functions, and only in certain positions. Moreover, some~\cite{StrzeboskiExpLogArctan,StrzebonskiTame, ReductionTranscDecProb} depend on a currently unproved conjecture (Schanuel's conjecture). Examples of such fragments are exp–log–arctan functions~\cite{StrzeboskiExpLogArctan, MCCALLUM201216}, tame elementary functions~\cite{StrzebonskiTame}, poly-powers~\cite{RealRootIsolPolyPower}, mixed trigonometric-polynomials~\cite{RootIsolMixedTrigPol}, and trigonometric extensions~\cite{ReductionTranscDecProb}. Recently, by leveraging such real root isolation algorithms, decision procedures have been shown for the theory of univariate mixed trigonometric-polynomials~\cite{DecidingMixedTrigPol} and for trigonometric extensions~\cite{ReductionTranscDecProb}.
Unlike these techniques,  our method tackles all of \nta, without any syntactical restrictions.

We are only aware of two approaches that extend verification techniques for square systems of equations to proving satisfiability of quantifier-free non-linear arithmetic~\cite{raSAT,ATVApaper}, one~\cite{raSAT} being restricted to the polynomial case, and the other one also being able to handle transcendental function symbols. Neither approach  is formulated in the form of certificate search. However, both could be interpreted as such, and both could be extended to return a certificate. The present paper actually does this for the second approach~\cite{ATVApaper}, and demonstrates that this does not only ease the independent verification of results, but also allows the systematic design of search techniques that result in significant efficiency improvements. 

An alternative approach is to relax the notation of satisfiability, for example using the notion of $\delta$-satisfiability~\cite{dreal,ksmt2}, that does \emph{not} guarantee that the given formula is satisfiable, but only that the formula is not too far away from a satisfiable one, for a suitable formalization of the notion of ``not too far away''. Another strategy is to return candidate solutions in the form of bounds that guarantee that certain efforts to prove unsatisfiability within those bounds fail~\cite{iSAT3}.


\section{Conclusions}
\label{sec:conclusions}
We introduced a form of satisfiability certificate for \smtnta and formulated the satisfiability verification problem as the problem of searching for such a certificate. We showed how to perform this search in a systematic fashion introducing new and efficient search techniques, and provided a theoretical classification providing insight into the possibilities and restrictions of such an approach. Computational experiments document that the resulting method is able to prove satisfiability of a substantially higher number of benchmark problems than existing methods. This suggests future work on the  integration of such search methods into the deduction machinery of current SMT solvers.

While in the unsatisfiable case~\cite{Barbosa:23}, proof certificates are already produced by some current SMT solvers,  and there has already been a standardization effort concerning proof formats and software infrastructure, this is still a topic for future work in the case of satisfiability of \smtnta. Finally, the extension of such techniques to even richer theories (e.g., reasoning not only about real numbers, but about real-valued functions~\cite{Ratschan:23}), is an interesting topic for future research.


\section*{Acknowledgments}\label{s:ack}
The work of Stefan Ratschan was supported by
the project 21-09458S of the Czech Science Foundation GA ČR
and institutional support RVO:67985807.

\bibliographystyle{plain}
\bibliography{refs}

\end{document}